\documentclass[letter]{aa} 

\def\oiiia/{{[$\ion{\rm O}{III}$]\,5007\AA\xspace}}
\def\oiiib/{{[$\ion{\rm O}{III}$]\,4959\AA\xspace}}
\def\niia/{{[$\ion{\rm N}{II}$]\,6548\AA\xspace}}
\def\niib/{{[$\ion{\rm N}{II}$]\,6584\AA\xspace}}
\def\siia/{{[$\ion{\rm S}{II}$]\,6717\AA\xspace}}
\def\siib/{{[$\ion{\rm S}{II}$]\,6731\AA\xspace}}
\def\sii/{{$[\ion{S}{II}$]6717,6731\AA\xspace}}
\def\heii/{{$\ion{\rm He}{II}$\,4686\AA\xspace}}
\def\ha/{{$H_{\alpha}$}}
\def\hb/{{$H_{\beta}$}}
\def\hg/{{$H_{\gamma}$}}

\newcommand*{\arcsecf}{\hbox{$.\!\!^{\prime\prime}$}}
\usepackage{graphicx}
\usepackage{txfonts}
\usepackage{xspace}
\usepackage{multirow}
\usepackage{booktabs}
\usepackage{float}
\defcitealias{CF00}{CF00}
\defcitealias{Calzetti2000}{C00}

\begin{document}

   \title{Spectroscopic confirmation of a dust-obscured, possibly metal-rich dwarf galaxy at $z\sim5$}
   \titlerunning{A dusty dwarf at $z\sim5$}
   \authorrunning{Bisigello et al.}

\author{Bisigello, L.$^{1,2}, $\thanks{\email{laura.bisigello@inaf.it}}, Gandolfi, G.$^{2}$, Feltre, A.$^{3}$, Arrabal Haro, P.$^{4,5}$ $\thanks{NASA Postdoctoral Fellow}$, Calabrò,A.$^{6}$, Cleri, N. J.$^{7,8,9}$, Costantin, L.$^{10}$, Girardi, G.$^{2,11}$, Giulietti, M.$^{1}$, Grazian, A.$^{11}$, Gruppioni, C.$^{12}$, Hathi, N. P.$^{13}$ , Holwerda, B.W.$^{14}$, Llerena, M.$^{15}$, Lucas, R.A.$^{13}$, Pacucci, F.$^{16,17}$, Prandoni, I.$^{1}$, Rodighiero, G.$^{2}$, Seillé, L. -M.$^{18}$, Wilkins, S. M.$^{19,20}$, Bagley, M.$^{21}$, Dickinson., M.$^{4}$, Finkelstein, S.L.$^{21,22}$, Kartaltepe, J.$^{23}$, Koekemoer, A. M.$^{13}$, Papovich, C.$^{24,25}$, Pirzkal, N.$^{26}$} 

\institute{$^{1}$ INAF, Istituto di Radioastronomia, Via Piero Gobetti 101, 40129 Bologna, Italy\\
$^{2}$ Dipartimento di Fisica e Astronomia "G. Galilei", Universit\`a di Padova, Via Marzolo 8, 35131 Padova, Italy\\
$^{3}$ INAF-Osservatorio Astrofisico di Arcetri, Largo E. Fermi 5, 50125, Firenze, Italy\\
$^{4}$ NSF's National Optical-Infrared Astronomy Research Laboratory, 950 N. Cherry Ave., Tucson, AZ 85719, USA\\
NASA Postdoctoral Fellow\\
$^{5}$ Astrophysics Science Division, NASA Goddard Space Flight Center, 8800 Greenbelt Rd, Greenbelt, MD 20771, USA\\
$^{6}$ INAF - Osservatorio Astronomico di Roma, via di Frascati 33, 00078 Monte Porzio Catone, Italy\\
$^{7}$Department of Astronomy and Astrophysics, The Pennsylvania State University, University Park, PA 16802, USA \\
$^{8}$Institute for Computational and Data Sciences, The Pennsylvania State University, University Park, PA 16802, USA \\
$^{9}$Institute for Gravitation and the Cosmos, The Pennsylvania State University, University Park, PA 16802, USA \\
$^{10}$Centro de Astrobiolog\'ia (CAB), CSIC-INTA, Ctra de Ajalvir km 4, Torrej\'on de Ardoz, 28850, Madrid, Spain\\
$^{11}$ INAF--Osservatorio Astronomico di Padova, Vicolo dell'Osservatorio 5, I-35122, Padova, Italy \\
$^{12}$ INAF-Osservatorio di Astrofisica e Scienza dello Spazio, via Gobetti 93/3, I-40129, Bologna, Italy\\
$^{13}$ Space Telescope Science Institute, 3700 San Martin Drive,Baltimore, MD 21218, USA\\
$^{14}$ Physics \& Astronomy Department, University of Louisville, 40292 KY, Louisville, USA\\
$^{15}$INAF - Osservatorio Astronomico di Roma, via di Frascati 33, 00078 Monte Porzio Catone, Italy \\
$^{16}$Center for Astrophysics $\vert$ Harvard \& Smithsonian, 60 Garden St, Cambridge, MA 02138, USA\\
$^{17}$Black Hole Initiative, Harvard University, 20 Garden St, Cambridge, MA 02138, USA\\
$^{18}$Aix Marseille Univ, CNRS, CNES, LAM, Marseille, France\\
$^{19}$ Astronomy Centre, University of Sussex, Falmer, Brighton BN1 9QH, UK\\
$^{20}$ Institute of Space Sciences and Astronomy, University of Malta, Msida MSD 2080, Malta \\
$^{21}$ Department of Astronomy, The University of Texas at Austin, Austin, TX, USA\\
$^{22}$Cosmic Frontier Center, The University of Texas at Austin, Austin, TX, USA \\
$^{23}$Laboratory for Multiwavelength Astrophysics, School of Physics and Astronomy, Rochester Institute of Technology, 84 Lomb Memorial Drive, Rochester, NY 14623, USA\\
$^{24}$Department of Physics and Astronomy, Texas A\&M University, College Station, TX, 77843-4242 USA \\
$^{25}$George P.\ and Cynthia Woods Mitchell Institute for Fundamental Physics and Astronomy, Texas A\&M University, College Station, TX, 77843-4242 USA\\
$^{26}$ESA/AURA Space Telescope Science Institute\\
}

   \date{Received ; accepted }

 
  \abstract
 {We present the first spectroscopic confirmation of a dust-obscured dwarf galaxy, CEERS-14821. 
 The analysis was performed by combining JWST NIRCam broadband photometry and NIRSpec/PRISM spectroscopic data. From the detection of multiple rest-frame optical lines, we derive that CEERS-14821 is located at $z=4.883\pm0.003$. Moreover, from a secure detection of the \ha/ and \hb/, we derive that the galaxy has a dust extinction ranging from $A_{V}=2.2_{-0.6}^{+0.5}$ to $A_{V}=3.4_{-0.9}^{+0.7}$, depending on the assumed reddening law. This value is extremely large given that we estimated a low stellar mass, that is, $\rm log_{10}(M/M_{\odot})=8.17_{-0.04}^{+0.05}$ or $\rm log_{10}(M/M_{\odot})=8.65_{-0.05}^{+0.06}$, based on two different dust extinction laws. Moreover, the combination of different metallicity tracers and the spectro-photometric fit suggests that the galaxy may also be metal-rich, with $\rm 12+log_{10}(O/H)>8.3$, but a low metallicity value cannot be totally ruled out. The high metallicity value would be above the expectation based on the mass-metallicity relation. Both metallicity estimations are above the expectations based on the fundamental mass-metallicity relation since CEERS-14821 is going through a burst of star formation. The constraints on a possible active galactic nucleus presence are limited and loose, but they point towards a possible non-dominant contribution ($f_{AGN}<0.5$ with respect to the total dust luminosity). Based on the rest-frame optical images, this source has a size compatible with galaxies of similar stellar masses and at similar redshifts. Finally, CEERS-14821 may be part of a larger galaxy overdensity, but there are no other galaxies closely interacting with it  (within 30 Mpc). }

   \keywords{}

   \maketitle
%
\begin{nolinenumbers}

\section{Introduction}
Only $1\%$ of the total mass of the interstellar medium (ISM) is composed of dust, which is key to the formation and cooling of the molecular gas and the formation of new stars \citep[for a review, see][]{Draine2003}. The presence of dust can also profoundly impact the derivation of physical properties and our understanding of galaxy evolution as it absorbs ultraviolet (UV) and optical radiation and re-emits it at infrared (IR) wavelengths. \par

Several studies have focused on dust-obscured objects and their dust mass \citep[e.g.][]{Gruppioni2013,Bethermin2015,Zavala2021,Gentile2024,Traina2024,Traina2024b} and on objects so deeply obscured to be extremely faint or completely undetected in the rest-frame optical and UV \citep[e.g.][]{Rodighiero2007,Simpson2014,Franco2018,Wang2019,Talia2021,Gruppioni2020,Enia2022,Gentile2024b}. To our understanding, dust-obscured galaxies dominate the cosmic star formation budget at least out to $z=4-5$ \citep[][Traina et al. in prep.]{Zavala2017,Magnelli2024} and the massive end of the stellar mass function \citep{Rodighiero2007,Caputi2015,WangT2016,Wang2019}. However, all these studies were hampered by the faintness of these objects in the optical and the limitations of IR instruments, which impacts the robustness of derived physical parameters and, as a consequence, our understanding of these galaxies. The \textit{James Webb} Space Telescope (JWST) is finally allowing us to observe the stellar continuum of these objects with high angular resolution and sensitivity, enabling the study of dust-obscured objects to higher redshifts and lower stellar masses than previously feasible \citep[e.g.][]{PerezGonzalez2023,Rodighiero2023,Xiao2023,Wang2024,Williams2024}. \par

In this respect, the studies by \citet{Bisigello2023b} and \citet{Rodighiero2023} indicate the presence of a possible population of dust-obscured galaxies with low stellar masses. These objects were unexpected given that the dust extinction ($A_{v}$) of star-forming galaxies has a positive correlation with stellar mass ($M_{*}$), with more massive galaxies being generally dustier, whereas low-mass galaxies (with $M_{*}<10^{8.5}\,\rm M_{\odot}$) typically have a negligible dust content \citep{Pannella2015,McLure2018,Shapley2023,Liu2024}. In fact, dust is produced by mass loss from asymptotic giant branch stars and supernova ejecta \citep[e.g.][]{Gall2011,Sarangi2018}, so an increase in star formation results in an increase in stellar mass and dust mass. Moreover, stellar-feedback-driven outflows can efficiently expel metal-enriched gas from low-mass galaxies, given their low gravitational potential \citep{Dayal2013}. In addition, dust in low-mass galaxies has a much larger scale height and thus appears more diffuse than in massive ones, resulting in an overall lower dust extinction \citep{Dalcanton2004}.
Thus, abundant dust extinction in low-mass systems is unexpected and could have a non-negligible impact on the overall cosmic star formation budget. The results presented in \citet{Bisigello2023b} and \citet{Rodighiero2023}, based on photometric data only, must therefore be confirmed with robust spectroscopic measurements.\par

In this Letter we present the first spectroscopic confirmation of a high dust extinction in a low-mass galaxy at $z=4.88$, using JWST NIRSpec (Near-Infrared Spectrograph) PRISM data. We present the observations in Sect. \ref{sec:data}, the analysis in Sect. \ref{sec:analysis}, and discuss the main findings and their interpretation in Sect.
\ref{sec:discussion}. We summarise our results in Sect. \ref{sec:summary}. Throughout the paper, we assume a $\Lambda$ cold dark matter cosmology with $H_0=70\,{\rm km}\,{\rm s}^{-1}\,{\rm Mpc}^{-1} $, $\Omega_{\rm m}=0.27$, and $\Omega_\Lambda=0.73$ and a Kroupa initial mass function \citep[][]{Kroupa2001}, and all magnitudes are in the AB system \citep{Oke1983}.

\section{Data} \label{sec:data}
\subsection{Photometry}
We used the source catalogue from \citet{Finkelstein2024}, which is based on the v0.6 release of the imaging data \citep{Bagley2023} of the Cosmic Evolution Early Release Science Survey (CEERS; P.I. Finkelstein) covering the Extended Groth Strip. These data include observations in seven NIRCam broad and medium bands (F115W, F150W, F200W, F277W, F356W, F444W, and F410M). In addition, there are deep \textit{Hubble} Space Telescope (HST) observations \citep{Grogin2011,Koekemoer2011} in six filters (ACS/WFC F606W, F814W, and WFC3/IR F105W, F125W, F140W, and F160W).

As explained in detail in \citet{Finkelstein2024}, photometric fluxes were derived with the Source Extractor \citep[v2.25.0][]{Bertin1996} in 
dual-image mode, using as detection images the inverse variance weighted images of the F277W and F356W bands at the native resolution. The fluxes were measured in each HST and JWST image on small elliptical Kron apertures, corrected for any missing flux by comparing them with larger Kron apertures. We refer to \citet{Finkelstein2023a} and \citet{Finkelstein2024} for further details.  

CEERS-14821 ($\rm RA=214.917994$, $\rm Dec=52.937245$) is part of a larger sample of highly extincted low-mass (HELM) galaxy candidates ($\sim4000$ objects; Bisigello et al. in prep.). The selection was performed considering the multi-parameter probability distribution of $M_{*}$ and $A_V$ obtained by fitting the photometric data using the spectral energy distribution (SED) fitting code BAGPIPES \citep[Bayesian Analysis of Galaxies for Physical Inference and Parameter EStimation;][]{Carnall2018} with different configurations. The sample notably includes objects with a more than 68\% probability of being either a galaxy with a low stellar mass and $A_{v}>1$ or a galaxy with a higher mass and a dust extinction above the expectations:
\begin{align}\label{eq:selection}
    &(\rm log_{10}(M_{*}/M_{\odot})\leq8.5) \land (A_{v}>1) \;\lor\\
    &(\rm log_{10}(M_{*}/M_{\odot})>8.5) \land (A_{v}>1.6\,\rm log_{10}(M/M_{\odot})-12.6).\nonumber
\end{align}
Given the finding of no redshift evolution in the $A_{v}-M_{*}$ relation at least out to $z=6.5$ \citep{Shapley2023}, we chose to not evolve the selection either. CEERS-14821 had a 96.8\%  probability of being inside the selection and is the only object with available spectroscopic data. For further details on the SED fitting and selection criterion, we refer to Bisigello et al. (in prep). In Appendix \ref{sec:images} we report the images and the photometric fluxes. CEERS-14821 is robustly detected in the four filters at the longest wavelengths, and it is spatially resolved in two of them (F277W and F356W).

\subsection{Spectroscopy}
NIRSpec Micro Shutter Array \citep[MSA;][]{Ferruit2022} spectra of CEERS-14821 were taken as part of programme DD-2750 \citep[P.I. Arrabal Haro;][]{ArrabalHaro2023b}, in which the source (MSA ID D2015) was observed in three visits each with three integrations of nine or ten groups in NRSIRS2 readout mode, for a total of 18387 s ($\sim$5.1 h) of combined integration time. The configuration employed three-shutter slitlets to enable a three-point nodding pattern and made use of the NIRSpec prism, providing a complete wavelength coverage from 0.6 to 5.3 $\mu$m (corresponding to the 0.1-0.9 $\mu$m rest frame at $z=4.883$) at a variable spectral resolution ($R\approx30-300$ from blue to red).
The data reduction followed the process described in \citet{ArrabalHaro2023a}, with the additional improvements mentioned below that will be further detailed in Arrabal Haro (in prep.).
The Calibration Reference Data System (CRDS) mapping 1227 was used with the STScI calibration pipeline\footnote{\url{https://jwst-pipeline.readthedocs.io/en/latest/index.html}} \citep{Bushouse2024} version 1.14.0. During the reduction process, a custom correction was applied to the \texttt{FAST\_VARIATION} extension of the NIRSpec \texttt{FFLAT}, which would otherwise result in an unrealistically low signal-to-noise ratio (S/N) in the 2.5-3 $\mu$m range.
A default pipeline \texttt{pathloss} correction was adopted to partly account for slit losses, and a further calibration via a comparison to the NIRCam photometry was carried out to correct for this effect (see Sect.~\ref{sec:linefit}).
The resulting spectrum from each independent visit after a nodding combination was inspected to identify and remove any remaining hot pixels or image defects, making use of the Mosviz visualisation tool\footnote{\url{https://jdaviz.readthedocs.io/en/latest/mosviz/index.html}} \citep{JDADF2023}. After that, the masked independent visits were combined with an inverse variance weighting. It is important to note that all the visits in the DD-2750 programme employed the same pointing and MSA configuration. Thus, the target source was at the exact same location within its shutter in all the visits, which facilitated their combination as no spatial or spectral resampling was required in the process.
The final 1D spectrum presented in Fig.~\ref{fig:spectrum} was extracted from the combined 2D spectrum following the optimal extraction prescriptions in \cite{Horne1986}.

\section{Analysis}\label{sec:analysis}
\subsection{Emission line fitting}\label{sec:linefit}

The flux of each nebular emission line was derived by fitting a Gaussian function to the spectrum, after removing the stellar continuum and taking the slit loss into account. The fit was performed simultaneously when the lines were blended. In particular, \ha/, \niia/, and \niib/ are severely blended, making their flux estimates highly covariant and impacting our results (see Appendix \ref{sec:lfmore}). We give all the details of the fit in Appendix \ref{sec:lfmore}, and we report the measured line fluxes in Table \ref{tab:lines}.

From the \ha/ / \hb/ ratio (i.e. $5.42\pm0.85$), assuming a case-B recombination at $T = 1.5\times10^4\,K$ (i.e. $H_{\alpha} / H_{\beta}=2.86$) and the \citet[][hereafter C00]{Calzetti2000} reddening law, we derived a dust extinction in the V band of $A_{v,neb}=2.2_{-0.6}^{+0.5}$. The value is even larger if we consider the \citet[][hereafter CF00]{CF00}  reddening law with a slope of -0.7, which results in $A_{v,neb}=3.4_{-0.9}^{+0.7}$, and it remains above $A_{v,neb}=1$ even if we consider extremely steep dust extinction slopes (Appendix \ref{sec:slope}). The S/N of the $H_{\gamma} / H_{\beta}$ ratio is unfortunately too low to give significant results, and moreover, the $H_{\gamma}$ is blended with the $[\ion{O}{III}]4663\AA$ line. We also used the dust-corrected \ha/ flux, which is sensitive to recent star formation  ($<10Myr$), to extract the star formation rate (SFR) using the relation from \citet{Kennicutt2012}. We estimate a $\rm SFR=41_{-17}^{+17}\,M_{\odot}/yr$ and $\rm SFR=114_{-64}^{+88}\,M_{\odot}/yr$ based on the \citetalias{Calzetti2000} and \citetalias{CF00} reddening laws, respectively. Consistent SFR estimates are derived from the dust-corrected \hb/ line.

We used the $[\ion{\rm O}{III}]\,5007\AA/H_{\beta}$ ratio (R3) to estimate the gas-phase metallicity, obtaining $\rm 12+log_{10}(O/H)=8.38_{-0.04}^{+0.04}$ with the \citet{Curti2024} relation. We derived a consistent estimate using the \citet{Nakajima2022} relation. We note that R3 depends on the ionisation parameter of the gas and is a double-value calibrator. The low-metallicity estimation is $\rm 12+log_{10}(O/H)=7.30_{-0.05}^{+0.06}$. The lower limit on the $[\ion{\rm O}{III}]\,5007\AA/\ion{\rm O}{II}]\,3727\AA$ ratio (O32), which is a metallicity estimation, is $\rm 12+log_{10}(O/H)<8.46$; this is not sufficient to discriminate between the two R3 estimations. The \ha/ and \niib/ lines, whose ratio (N2) can also be used to estimate the gas-phase metallicity, are severely blended. Taking this ratio with extreme caution and using the \citet{Curti2024} relation, or similarly the \citet{Pettini2004} and \citet{Nakajima2022} relations, we obtain a higher metallicity, $12+log_{10}(O/H)=8.77_{-0.17}^{+0.11}$. The $5\sigma$ upper limit on the $[\ion{\rm S}{II}]\,6717,6731\AA/H_{\alpha}$ (S2), again based on  the \citet{Curti2024} relation, gives an upper limit on the gas-phase metallicity of $\rm 12+log_{10}(O/H)<8.54$, which is consistent with both R3 estimates. Finally, if we estimate the $H_{\gamma}$ line from the $H_{beta}$ line, assuming case-B recombination, we obtain an estimate for the $[\ion{\rm O}{III}]\,4636\AA$ flux, as the two lines are blended. The $[\ion{\rm O}{III}]\,4636\AA$ line, combined with the $[\ion{\rm O}{III}]\,4859\AA$ and $[\ion{\rm O}{III}]\,5007\AA$ lines, can then be used to estimate the electron temperature of the high-ionisation region ($t_{3}$; \citealt{PerezMontero2017}). The electron temperature of the low-ionisation region ($t_{2}$) can then be estimated using the \citet{Pilyugin2009} relation. These temperatures, combined with the ionic abundances of oxygen \citep{PerezMontero2017}, can then be used to directly estimate the metallicity. We propagated the uncertainties on the $[\ion{O}{III}]5007\AA$ and $H_{\beta}$ lines, as well as that on the dust extinction, deriving a value of $12+log_{10}(O/H)=6.88_{-0.33}^{+0.67}$, which is inconsistent with the R3 high-metallicity estimation. However, it is necessary to understand that an underestimation of the $[\ion{O}{III}]4363\AA$ auroral line, due to stellar absorption in the continuum or an underestimation of the dust extinction, would result in an underestimation of the metallicity. On the other hand, we only have upper limits for the $[\ion{O}{II}]3727\AA$ emission line, which indicates that the overall metallicity could be overestimated. We discuss these results further in Sect. \ref{sec:discussion} and report the derived physical properties in Table \ref{tab:param}.

\subsection{Physical size}
To estimate the physical size of this galaxy, we derived the radial profile in the two filters where the galaxy is detected and resolved. We then fitted it with a Sérsic profile \citep{Sersic1968} convolved with the JWST point spread function (PSF) taken from WebbPSF V1.1.0 \citep{Perrin2012}. 
We derived an effective radius of $ Re=1.2_{-0.5}^{+1.3}\,kpc$ (corresponding to $0\arcsecf19_{-0.07}^{+0.20}$) in the F277W filter and $ Re=0.7_{-0.1}^{+0.2}\,kpc$ (corresponding to $0\arcsecf10_{-0.01}^{+0.03}$) in the F356W filter. Cutouts of CEERS-14821 in all HST and JWST filters are shown in Fig. \ref{fig:cutouts}, and we show the light profiles in Fig. \ref{fig:profiles}.
 
\subsection{Spectro-photometric SED fitting} \label{sec:SEDfit}

We performed a combined spectro-photometric fit using the 13 HST and NIRCam filter observations and the NIRSpec spectrum, including both nebular lines and stellar continuum. The spectrum was normalised using a second-order polynomial (see Sect. \ref{sec:linefit}). To perform the SED fitting, we used the BAGPIPES code, fixing the redshift to the measured spectroscopic value. We considered two different dust extinction laws (\citetalias{Calzetti2000,CF00}) and used the values derived from the Balmer decrement as priors for $A_{V}$, rescaling them to take the differential attenuation into account. In addition, in an attempt to verify or rule out the presence of an active galactic nucleus (AGN), we  not only used several line diagnostics (see Sect. \ref{sec:discussion}), but we also fitted the available photometry with the \texttt{CIGALE} code \citep{Boquien2019} using similar parameters and priors. In Appendix D we give more information on the fits and the configurations used. We consider the BAGPIPES results as the fiducial ones, but CEERS-14821 is consistent with being a dusty dwarf even with \texttt{CIGALE}.

\section{Discussion}\label{sec:discussion} 

\begin{table}[]
    \caption{Derived physical properties of CEERS-14821.}
    \centering
    \begin{tabular}{c|c}
    Physical parameter & Estimated value \\
    \hline
       $z_{spec}$$^{a}$  &  $4.883\pm0.003$\\
       \multirow{2}{*}{$\rm 12+log_{10}(O/H)$ (R3)}$^{a,b}$ & $8.38_{-0.04}^{+0.04}$ \\ 
       & $7.30_{-0.05}^{+0.06}$ \\
        $R_{e}\,[kpc]^{c}$ & $0.7_{-0.1}^{+0.2}$ \\
       \hline
       $A_{v,neb}$ - \citetalias{Calzetti2000}$^{a}$ & $2.2_{-0.6}^{+0.5}$ \\
       $\rm log_{10}(M/M_{\odot})^{d}$ & $8.17_{-0.04}^{+0.05}$ \\
        SFR $[M_{\odot}/yr]$ (SED,$<100\,Myr$)$^{d}$ & $4^{+3}_{-2}$ \\
        SFR $[M_{\odot}/yr]$ (SED,$<10\,Myr$)$^{d}$ & $16^{+2}_{-2}$ \\
        SFR $[M_{\odot}/yr]$ (\ha/,$<10\,Myr$)$^{a}$ & $41_{-17}^{+17}$ \\
        \hline
       $A_{v,neb}$ - \citetalias{CF00}$^{a}$ & $3.4_{-0.9}^{+0.7}$ \\
       $\rm log_{10}(M/M_{\odot})^{c}$ & $8.65_{-0.05}^{+0.06}$ \\ 
        SFR $[M_{\odot}/yr]$ (SED,$<100\,Myr$)$^{d}$ & $5^{+3}_{-2}$ \\ 
        SFR $[M_{\odot}/yr]$ (SED,$<10\,Myr$)$^{d}$ & $48_{-6}^{+7}$ \\ 
        SFR $[M_{\odot}/yr]$ (\ha/,$<10\,Myr$)$^{a}$ & $114_{-64}^{+88}$ \\
        
    \end{tabular}
    \label{tab:param}
    \tablefoot{$^{a}$ Derived from the spectrum. $^{b}$ The high metallicity value is slightly preferred considering the N2 and SED fitting results. $^{c}$ Derived from the F356W image. $^{d}$ Derived from the spectro-photometric fit.
    For some parameters, we report the estimates derived using the V-band dust extinction and assuming two different extinction laws \citepalias{Calzetti2000,CF00}. }
\end{table}

In Table \ref{tab:param} we summarise the physical properties of CEERS-14821 derived from the images, the line fluxes, and the BAGPIPES spectro-photometric SED fitting. 

The first surprising result is the high $A_{V}$ measured given the stellar mass of this object (Fig. \ref{fig:MAV}). We can take as reference the continuum $A_{1600,cont}-M_{*}$ relation from \citet{McLure2018}. This relation was derived considering galaxies at $2 < z < 3$ and with $9.25\leq log (M_{*}/M_{\odot}) \leq 10.75$, but it has been shown to be consistent with other works at similar or higher redshifts \citep{Shapley2022,Shapley2023}. Using this $A_{1600,cont}-M_{*}$ relation, we would expect a dust extinction at 1600$\AA$ of $A_{1600,cont}\leq0.8$ considering our uncertainties on the stellar mass. This would correspond to $A_{V,cont}\leq0.33$ with the \citetalias{CF00} reddening law and $A_{V,cont}=0$ with the \citetalias{Calzetti2000} reddening law. Even after taking into account the differential attenuation of the nebular emission with respect to the stellar UV-optical continuum (known as the f-factor), which can be a factor of two to three \citep[e.g.][]{Calzetti1994,CF00,Kreckel2013,Talia2015,Buat2018,RodriguezMunoz2022}, it is clear that our source is much more dust extinguished than expected given its stellar mass. Similarly, in the First Light And Reionization Epoch Simulations (FLARES), the majority of galaxies ($84\%$) with $\rm M_{*}=10^{8}\,M_{\odot}$ at $z=5$ have $A_{1500}<0.4$ with a few extending up to $A_{1500}=1$ \citep{Vijayan2021}, which is still too low compared with CEERS-14821. The peculiar dusty nature of CEERS-14821 is also supported by the fact that sources with similar stellar masses are expected to have an extremely blue UV-continuum slope, such as $\beta=-2$, but the UV continuum is so obscured in CEERS-14821 that it is undetected (i.e. $f_{F814W}=6.41\pm5.78\,nJy$). 

\begin{figure*}
    \centering
    \includegraphics[width=0.9\linewidth]{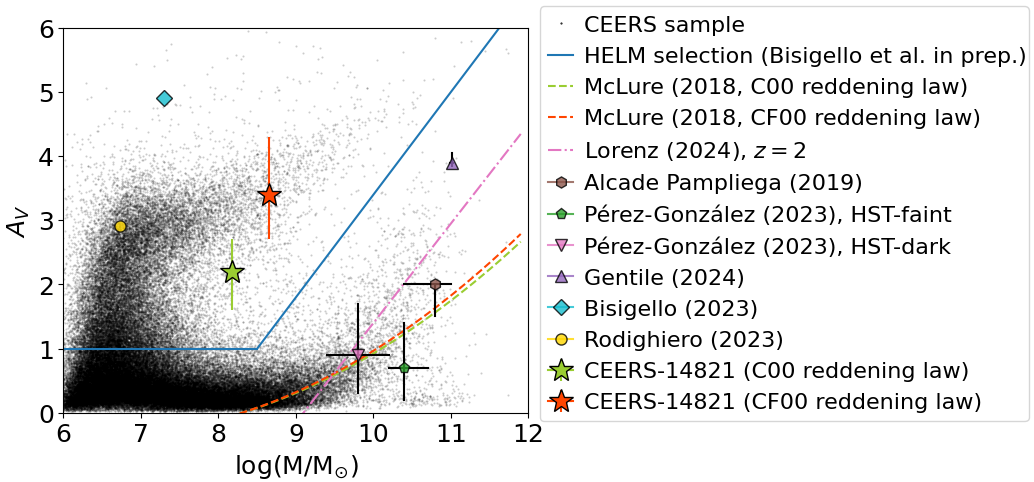}
    \caption{Stellar mass vs dust extinction. We compare CEERS-14821 (coloured stars) with the complete CEERS sample (black dots; Bisigello et al. in prep.) at all redshifts. We also show the median values of other samples of dusty objects \citep{AlcaldePampliega2019,PerezGonzalez2023,Gentile2024} and low-mass dusty galaxies \citep{Rodighiero2023,Bisigello2023} from the literature. We report the selection criteria for highly extincted low-mass  galaxies in Eq. \ref{eq:selection} (solid blue line; Bisigello et al. in prep.) and the relation between stellar mass and $A_{V}$ from the literature \citep[dashed coloured lines;][]{McLure2018,Lorenz2024}.}
    \label{fig:MAV}
\end{figure*}

Given that dust content is closely related to gas metallicity, an obvious step is to look at the mass-metallicity relation \citep[MZR;][]{Tremonti2004,Maiolino2019}. If we consider the relation at $z=3-6$ from \citet{Curti2024}, based on JWST PRISM data, we would expect to observe a gas-phase metallicity around $12+log_{10}(O/H)=7.7-7.9$ given the stellar mass (Fig. \ref{fig:MZR}). As mentioned previously, the gas-phase metallicity derived using the R3 method, which is the most robust given the available data, gives two possible results, one above and one below the expectation based on the MZR. Considering the results from the N2 tracer, which, however, is severely affected by blending, and from the spectro-photometric SED fitting, the high-metallicity result would be preferred. This result is, however, slightly in tension with the direct method estimation, which is limited by the blending between $H_{\gamma}$ and $[\ion{O}{III}]4363\AA$. If we consider the expectation on the gas-phase metallicity from the fundamental metallicity relation \citep[FMR;][]{Mannucci2010}, including also the secondary dependence on the SFR, there is a clear discrepancy with both R3 estimations. We would expect $12+log_{10}(O/H)<6$ using the parametrisation from \citealt{Curti2020} and the SFR estimated based on the $H_{\alpha}$ line, as the metallicity decreases with the SFR at a fixed stellar mass. We can therefore postulate that the ISM of this galaxy has not only a relatively high dust content but possibly also a high gas-phase metallicity as compared to the bulk of the population with similar stellar masses and redshifts that follow the MZR and, particularly, the FMR.

\begin{figure}
    \centering
    \includegraphics[width=0.8\linewidth]{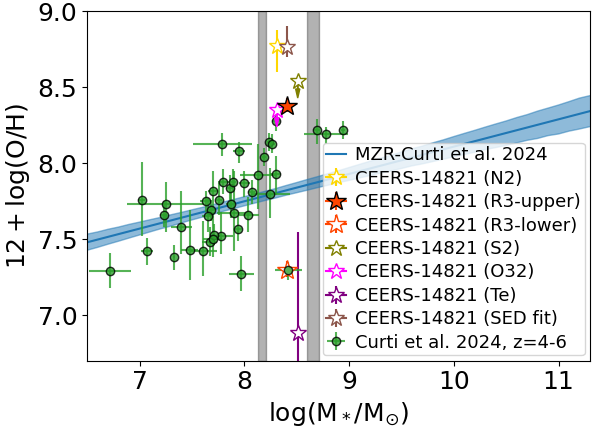}
    \caption{Comparison between the gas-phase metallicity of CEERS-14821 (coloured stars, slightly shifted horizontally for clarity) with the MZR (solid blue line) and other galaxies (green circles) at $z=4-6$ \citep{Curti2024}. Vertical grey shadows show the stellar mass of CEERS-14821 derived based on the two different dust extinction laws.}
    \label{fig:MZR}
\end{figure}

The star formation history derived from the spectro-photometric fit indicates that the galaxy is going through a burst of star formation (Fig. D.2). If we consider the main sequence of star-forming galaxies derived at $4\leq z<5$ by \citet{Santini2017}, we see that a galaxy with the stellar mass of CEERS-14821 is expected to have a $\rm SFR=0.6-2.6\,M_{\odot}/yr$. Similarly, if we consider the sequence derived by \citet{Cole2023} using a timescale for the SFR of 10 Myr, the galaxy is above the SFR expected based on its stellar mass, while it is in line with expectations if we consider the SFR over a longer timescale (i.e. 100 Myr). Therefore, CEERS-14821 is above the main sequence if we consider its current burst of star formation, but determining by how much strongly depends on the dust-extinction law considered.

Extrapolating the relation between stellar mass and size from \citet{Ward2024}, estimated for $z=3-5$ star-forming galaxies with $\rm log_{10}(M/M_{\odot})>9.5$, we derive a size of $R_{e}=0.6_{-0.1}^{+0.2}$ to $R_{e}=0.8_{-0.2}^{+0.1}\,kpc$, depending on the assumed stellar mass of CEERS-14821. This is consistent with the measured size. However, if we compare CEERS-14821 with the sample from \citet{Ono2024} at a similar redshift, which spans brighter galaxies, we derive that our source is larger than expected given the observed size-luminosity relation at $z=5$. CEERS-14821 is also larger than the general population of `little red dots' \citep[e.g.][]{Furtak2023,Baggen2024,Matthee2024,Guia2024}. However, the available observations trace the rest-frame optical and are sensitive to the stellar mass. We would need UV or IR data to trace the size of the star-forming regions, which indeed could be more compact \citep{Puglisi2019}. At the same time, the presence of dust can make galaxies appear larger since it smooths the light profile \citep{Roper2023,Costantin2023}. Therefore, current data indicate that compactness (a high $A_{V}$ due to geometry with an overall normal dust content) is likely not the cause of the high $A_{V}$.  However, more observations at higher angular resolution are necessary for a definitive answer. 

We also attempted to confirm or rule out the presence of an AGN using the \citet{BPT1981} AGN diagnostic (BPT) diagram. This is fundamental given that the presence of an AGN would impact our derivation of the SFR and the gas-phase metallicity, which is based on the assumption that the galaxy is dominated by star formation. With the separation described by \citet{Kewley2013} and \citet{Kauffmann2003}, we find that CEERS-14821 is in the composite region. We highlight that CEERS-14821 has an almost-solar metallicity, and therefore the BPT diagram should still be able to discriminate the presence of an AGN; this is not possible with other broad-line AGNs observed at similar redshifts that have very low metallicities \citep[e.g.][]{Ubler2023,Maiolino2023}. 

At the same time, based on the criterion from \citet{Kewley2006}, the upper limit on the \sii/ doublet places CEERS-14821 right on the separation line between star-forming and AGN-dominated systems. This indicates a significant star-forming contribution. The upper limit on the $[\ion{\rm O}{I}]\,6300\AA/H_{\alpha}$ ratio also implies that CEERS-14821 is dominated by star formation, again using the separation reported by \citet{Kewley2006}. The low spectral resolution of the spectrum prevents us from instead using the kinematics-excitation diagram \citep{Zhang2018} since the \oiiia/ line is unresolved. Similarly, the upper limit on the \heii/ line flux, unlike the measured \hb/ flux, is not stringent enough to rule out the presence of an AGN based on the relation from \citet{Shirazi2012}, as $log(\ion{He}{II}/H_{\beta})<-0.28$. Finally, the SED fitting, performed on the photometric data and including an AGN component, does not require a luminous AGN component, as the ratio of the AGN luminosity over the dust luminosity is $f_{AGN}=0.22\pm0.27$ (see Appendix D.2). Overall, there are no stringent and conclusive results, and we cannot completely rule out the presence of an AGN with the current data. However, the minimal evidence available points to, at most, a non-dominant contribution by an AGN.

Different works have reported a possible galaxy overdensity at z=4.88-4.91 in the CEERS field \citep{Naidu2022,ArrabalHaro2023b,deGraaff2024}, including 14 objects spectroscopically confirmed and a bright sub-millimetre galaxy with a consistent photometric redshift \citep{Zavala2023}. However, the closest spectroscopically confirmed object is $31\pm1\, Mpc$ from CEERS-14821. Therefore, none of the confirmed galaxies seem to be closely interacting with CEERS-14821, but further analyses are needed to understand if the high $A_{V}$ is driven by galaxy interactions.

\section{Summary}\label{sec:summary}
In this work we present the study of a $z=5$ dusty dwarf galaxy, CEERS-14821, based on broadband HST and JWST/NIRCam photometric data and JWST/NIRSpec spectroscopic data. We performed a detailed spectro-photometric analysis of this object, finding that it has a low stellar mass of $\rm log_{10}(M/M_{\odot})=8.17_{-0.04}^{+0.05}$ or $\rm log_{10}(M/M_{\odot})=8.65_{-0.05}^{+0.06}$, depending on the dust extinction law. Moreover, the galaxy also has a dust-rich and possibly metal-rich ISM. In particular, using the Balmer decrement, we measure $A_{V}=2.2_{-0.6}^{+0.5}$ and $A_{V}=3.4_{-0.9}^{+0.7}$ based on the \citet{Calzetti2000} and \citet{CF00} reddening laws, respectively. In both cases, the $A_{V}$ is well above the value expected given the low mass of CEERS-14821. At the same time, the gas-phase metallicity derived from the $[\ion{\rm O}{III}]\,5007\AA/H_{\beta}$ ratio, combined with the SED-fitting results, is $12+log_{10}(O/H)]=8.38\pm0.04$. However, based on the spectroscopic data alone, we are unable to totally rule out the secondary solution at $12+log_{10}(O/H)]=7.30_{-0.05}^{+0.06}$.  If the high-metallicity value is confirmed, it is well above the expectations based on the MZR. Both metallicity values are above the FMR given that, based on the spectro-photometric fit, CEERS-14821 seems to be going through a burst of star formation. 

The origin of the star formation burst and the richness of the ISM is still unclear. We confirm that CEERS-14821 has an optical size compatible with that of galaxies of similar masses and at similar redshifts, which rules out compactness as a reason for the large $A_{V}$.  We used several diagnostics in an attempt to confirm or rule out the presence of an AGN; while deeper spectroscopic data are needed to place stronger constraints, our current assessment points towards, at most, a non-dominant AGN contribution ($f_{AGN}<0.5$). At the same time, CEERS-14821 appears not to be interacting with other nearby objects. 

Spectroscopic data at higher spectral resolution are necessary to shed more light on the nature of this peculiar object, that is, to more robustly estimate the gas-phase metallicity and to look for a possible broad component in the hydrogen lines due to AGN activity.\\


\normalfont

\begin{acknowledgements}
LB acknowledges support from INAF under the Large Grant 2022 funding scheme (project "MeerKAT and LOFAR Team up: a Unique Radio Window on Galaxy/AGN co-Evolution”). This research made use of Photutils, an Astropy package for detection and photometry of astronomical sources 

\citep{photoutils}.

\end{acknowledgements}

%
\bibliographystyle{aa} 
\bibliography{main} 

\begin{thebibliography}{105}
\expandafter\ifx\csname natexlab\endcsname\relax\def\natexlab#1{#1}\fi

\bibitem[{{Alcalde Pampliega} {et~al.}(2019){Alcalde Pampliega}, {P{\'e}rez-Gonz{\'a}lez}, {Barro}, {Dom{\'\i}nguez S{\'a}nchez}, {Eliche-Moral}, {Cardiel}, {Hern{\'a}n-Caballero}, {Rodriguez-Mu{\~n}oz}, {S{\'a}nchez Bl{\'a}zquez}, \& {Esquej}}]{AlcaldePampliega2019}
{Alcalde Pampliega}, B., {P{\'e}rez-Gonz{\'a}lez}, P.~G., {Barro}, G., {et~al.} 2019, \apj, 876, 135

\bibitem[{{Arrabal Haro} {et~al.}(2023{\natexlab{a}}){Arrabal Haro}, {Dickinson}, {Finkelstein}, {Fujimoto}, {Fern{\'a}ndez}, {Kartaltepe}, {Jung}, {Cole}, {Burgarella}, {Chworowsky}, {Hutchison}, {Morales}, {Papovich}, {Simons}, {Amor{\'\i}n}, {Backhaus}, {Bagley}, {Bisigello}, {Calabr{\`o}}, {Castellano}, {Cleri}, {Dav{\'e}}, {Dekel}, {Ferguson}, {Fontana}, {Gawiser}, {Giavalisco}, {Harish}, {Hathi}, {Hirschmann}, {Holwerda}, {Huertas-Company}, {Koekemoer}, {Larson}, {Lucas}, {Mobasher}, {P{\'e}rez-Gonz{\'a}lez}, {Pirzkal}, {Rose}, {Santini}, {Trump}, {de la Vega}, {Wang}, {Weiner}, {Wilkins}, {Yang}, {Yung}, \& {Zavala}}]{ArrabalHaro2023a}
{Arrabal Haro}, P., {Dickinson}, M., {Finkelstein}, S.~L., {et~al.} 2023{\natexlab{a}}, \apjl, 951, L22

\bibitem[{{Arrabal Haro} {et~al.}(2023{\natexlab{b}}){Arrabal Haro}, {Dickinson}, {Finkelstein}, {Kartaltepe}, {Donnan}, {Burgarella}, {Carnall}, {Cullen}, {Dunlop}, {Fern{\'a}ndez}, {Fujimoto}, {Jung}, {Krips}, {Larson}, {Papovich}, {P{\'e}rez-Gonz{\'a}lez}, {Amor{\'\i}n}, {Bagley}, {Buat}, {Casey}, {Chworowsky}, {Cohen}, {Ferguson}, {Giavalisco}, {Huertas-Company}, {Hutchison}, {Kocevski}, {Koekemoer}, {Lucas}, {McLeod}, {McLure}, {Pirzkal}, {Seill{\'e}}, {Trump}, {Weiner}, {Wilkins}, \& {Zavala}}]{ArrabalHaro2023b}
{Arrabal Haro}, P., {Dickinson}, M., {Finkelstein}, S.~L., {et~al.} 2023{\natexlab{b}}, \nat, 622, 707

\bibitem[{{Baggen} {et~al.}(2024){Baggen}, {van Dokkum}, {Brammer}, {de Graaff}, {Franx}, {Greene}, {Labb{\'e}}, {Leja}, {Maseda}, {Nelson}, {Rix}, {Wang}, \& {Weibel}}]{Baggen2024}
{Baggen}, J. F.~W., {van Dokkum}, P., {Brammer}, G., {et~al.} 2024, arXiv e-prints, arXiv:2408.07745

\bibitem[{{Bagley} {et~al.}(2023){Bagley}, {Finkelstein}, {Koekemoer}, {Ferguson}, {Arrabal Haro}, {Dickinson}, {Kartaltepe}, {Papovich}, {P{\'e}rez-Gonz{\'a}lez}, {Pirzkal}, {Somerville}, {Willmer}, {Yang}, {Yung}, {Fontana}, {Grazian}, {Grogin}, {Hirschmann}, {Kewley}, {Kirkpatrick}, {Kocevski}, {Lotz}, {Medrano}, {Morales}, {Pentericci}, {Ravindranath}, {Trump}, {Wilkins}, {Calabr{\`o}}, {Cooper}, {Costantin}, {de la Vega}, {Hilbert}, {Hutchison}, {Larson}, {Lucas}, {McGrath}, {Ryan}, {Wang}, \& {Wuyts}}]{Bagley2023}
{Bagley}, M.~B., {Finkelstein}, S.~L., {Koekemoer}, A.~M., {et~al.} 2023, \apjl, 946, L12

\bibitem[{{Baldwin} {et~al.}(1981){Baldwin}, {Phillips}, \& {Terlevich}}]{BPT1981}
{Baldwin}, J.~A., {Phillips}, M.~M., \& {Terlevich}, R. 1981, \pasp, 93, 5

\bibitem[{{Bertin} \& {Arnouts}(1996)}]{Bertin1996}
{Bertin}, E. \& {Arnouts}, S. 1996, \aaps, 117, 393

\bibitem[{{B{\'e}thermin} {et~al.}(2015){B{\'e}thermin}, {Daddi}, {Magdis}, {Lagos}, {Sargent}, {Albrecht}, {Aussel}, {Bertoldi}, {Buat}, {Galametz}, {Heinis}, {Ilbert}, {Karim}, {Koekemoer}, {Lacey}, {Le Floc'h}, {Navarrete}, {Pannella}, {Schreiber}, {Smol{\v{c}}i{\'c}}, {Symeonidis}, \& {Viero}}]{Bethermin2015}
{B{\'e}thermin}, M., {Daddi}, E., {Magdis}, G., {et~al.} 2015, \aap, 573, A113

\bibitem[{{Bisigello} {et~al.}(2023){Bisigello}, {Gandolfi}, {Grazian}, {Rodighiero}, {Costantin}, {Cooray}, {Feltre}, {Gruppioni}, {Hathi}, {Holwerda}, {Koekemoer}, {Lucas}, {Newman}, {P{\'e}rez-Gonz{\'a}lez}, {Yung}, {de la Vega}, {Arrabal Haro}, {Bagley}, {Dickinson}, {Finkelstein}, {Kartaltepe}, {Papovich}, {Pirzkal}, \& {Wilkins}}]{Bisigello2023b}
{Bisigello}, L., {Gandolfi}, G., {Grazian}, A., {et~al.} 2023, \aap, 676, A76

\bibitem[{{Boquien} {et~al.}(2022){Boquien}, {Buat}, {Burgarella}, {Bardelli}, {B{\'e}thermin}, {Faisst}, {Ginolfi}, {Hathi}, {Jones}, {Koekemoer}, {Lemaux}, {Narayanan}, {Romano}, {Schaerer}, {Vergani}, {Zamorani}, \& {Zucca}}]{Boquien2022}
{Boquien}, M., {Buat}, V., {Burgarella}, D., {et~al.} 2022, \aap, 663, A50

\bibitem[{{Boquien} {et~al.}(2019){Boquien}, {Burgarella}, {Roehlly}, {Buat}, {Ciesla}, {Corre}, {Inoue}, \& {Salas}}]{Boquien2019}
{Boquien}, M., {Burgarella}, D., {Roehlly}, Y., {et~al.} 2019, \aap, 622, A103

\bibitem[{Bradley {et~al.}(2023)Bradley, Sip{\H o}cz, Robitaille, Tollerud, Vin{\'{\i}}cius, Deil, Barbary, Wilson, Busko, Donath, G{\"u}nther, Cara, Lim, Me{\ss}linger, Conseil, Bostroem, Droettboom, Bray, Bratholm, Barentsen, Craig, Rathi, Pascual, Perren, Georgiev, de~Val-Borro, Kerzendorf, Bach, Quint, \& Souchereau}]{photoutils}
Bradley, L., Sip{\H o}cz, B., Robitaille, T., {et~al.} 2023, astropy/photutils: 1.8.0

\bibitem[{{Bruzual} \& {Charlot}(2003)}]{Bruzual2003}
{Bruzual}, G. \& {Charlot}, S. 2003, \mnras, 344, 1000

\bibitem[{{Buat} {et~al.}(2018){Buat}, {Boquien}, {Ma{\l}ek}, {Corre}, {Salas}, {Roehlly}, {Shirley}, \& {Efstathiou}}]{Buat2018}
{Buat}, V., {Boquien}, M., {Ma{\l}ek}, K., {et~al.} 2018, \aap, 619, A135

\bibitem[{Bushouse {et~al.}(2024)Bushouse, Eisenhamer, Dencheva, Davies, Greenfield, Morrison, Hodge, Simon, Grumm, Droettboom, Slavich, Sosey, Pauly, Miller, Jedrzejewski, Hack, Davis, Crawford, Law, Gordon, Regan, Cara, MacDonald, Bradley, Shanahan, Jamieson, Teodoro, Williams, \& Pena-Guerrero}]{Bushouse2024}
Bushouse, H., Eisenhamer, J., Dencheva, N., {et~al.} 2024, JWST Calibration Pipeline

\bibitem[{{Calzetti} {et~al.}(2000){Calzetti}, {Armus}, {Bohlin}, {Kinney}, {Koornneef}, \& {Storchi-Bergmann}}]{Calzetti2000}
{Calzetti}, D., {Armus}, L., {Bohlin}, R.~C., {et~al.} 2000, \apj, 533, 682

\bibitem[{{Calzetti} {et~al.}(1994){Calzetti}, {Kinney}, \& {Storchi-Bergmann}}]{Calzetti1994}
{Calzetti}, D., {Kinney}, A.~L., \& {Storchi-Bergmann}, T. 1994, \apj, 429, 582

\bibitem[{{Caputi} {et~al.}(2015){Caputi}, {Ilbert}, {Laigle}, {McCracken}, {Le F{\`e}vre}, {Fynbo}, {Milvang-Jensen}, {Capak}, {Salvato}, \& {Taniguchi}}]{Caputi2015}
{Caputi}, K.~I., {Ilbert}, O., {Laigle}, C., {et~al.} 2015, \apj, 810, 73

\bibitem[{Carnall {et~al.}(2018)Carnall, McLure, Dunlop, \& Davé}]{Carnall2018}
Carnall, A.~C., McLure, R.~J., Dunlop, J.~S., \& Davé, R. 2018, MNRAS, 480, 4379

\bibitem[{{Charlot} \& {Fall}(2000)}]{CF00}
{Charlot}, S. \& {Fall}, S.~M. 2000, \apj, 539, 718

\bibitem[{{Chevallard} {et~al.}(2013){Chevallard}, {Charlot}, {Wandelt}, \& {Wild}}]{Chevallard2013}
{Chevallard}, J., {Charlot}, S., {Wandelt}, B., \& {Wild}, V. 2013, \mnras, 432, 2061

\bibitem[{{Cole} {et~al.}(2023){Cole}, {Papovich}, {Finkelstein}, {Bagley}, {Dickinson}, {Iyer}, {Yung}, {Ciesla}, {Amorin}, {Arrabal Haro}, {Bhatawdekar}, {Calabro}, {Cleri}, {de la Vega}, {Dekel}, {Endsley}, {Gawiser}, {Giavalisco}, {Hathi}, {Hirschmann}, {Holwerda}, {Kartaltepe}, {Koekemoer}, {Lucas}, {Mascia}, {Mobasher}, {Perez-Gonzalez}, {Rodighiero}, {Ronayne}, {Tachhella}, {Weiner}, \& {Wilkins}}]{Cole2023}
{Cole}, J.~W., {Papovich}, C., {Finkelstein}, S.~L., {et~al.} 2023, arXiv e-prints, arXiv:2312.10152

\bibitem[{{Costantin} {et~al.}(2023){Costantin}, {P{\'e}rez-Gonz{\'a}lez}, {Vega-Ferrero}, {Huertas-Company}, {Bisigello}, {Buitrago}, {Bagley}, {Cleri}, {Cooper}, {Finkelstein}, {Holwerda}, {Kartaltepe}, {Koekemoer}, {Nelson}, {Papovich}, {Pillepich}, {Pirzkal}, {Tacchella}, \& {Yung}}]{Costantin2023}
{Costantin}, L., {P{\'e}rez-Gonz{\'a}lez}, P.~G., {Vega-Ferrero}, J., {et~al.} 2023, \apj, 946, 71

\bibitem[{{Curti} {et~al.}(2024){Curti}, {Maiolino}, {Curtis-Lake}, {Chevallard}, {Carniani}, {D'Eugenio}, {Looser}, {Scholtz}, {Charlot}, {Cameron}, {{\"U}bler}, {Witstok}, {Boyett}, {Laseter}, {Sandles}, {Arribas}, {Bunker}, {Giardino}, {Maseda}, {Rawle}, {Rodr{\'\i}guez Del Pino}, {Smit}, {Willott}, {Eisenstein}, {Hausen}, {Johnson}, {Rieke}, {Robertson}, {Tacchella}, {Williams}, {Willmer}, {Baker}, {Bhatawdekar}, {Egami}, {Helton}, {Ji}, {Kumari}, {Perna}, {Shivaei}, \& {Sun}}]{Curti2024}
{Curti}, M., {Maiolino}, R., {Curtis-Lake}, E., {et~al.} 2024, \aap, 684, A75

\bibitem[{{Curti} {et~al.}(2020){Curti}, {Mannucci}, {Cresci}, \& {Maiolino}}]{Curti2020}
{Curti}, M., {Mannucci}, F., {Cresci}, G., \& {Maiolino}, R. 2020, \mnras, 491, 944

\bibitem[{{Dalcanton} {et~al.}(2004){Dalcanton}, {Yoachim}, \& {Bernstein}}]{Dalcanton2004}
{Dalcanton}, J.~J., {Yoachim}, P., \& {Bernstein}, R.~A. 2004, \apj, 608, 189

\bibitem[{{Dayal} {et~al.}(2013){Dayal}, {Ferrara}, \& {Dunlop}}]{Dayal2013}
{Dayal}, P., {Ferrara}, A., \& {Dunlop}, J.~S. 2013, \mnras, 430, 2891

\bibitem[{{de Graaff} {et~al.}(2024){de Graaff}, {Setton}, {Brammer}, {Cutler}, {Suess}, {Labbe}, {Leja}, {Weibel}, {Maseda}, {Whitaker}, {Bezanson}, {Boogaard}, {Cleri}, {De Lucia}, {Franx}, {Greene}, {Hirschmann}, {Matthee}, {McConachie}, {Naidu}, {Oesch}, {Price}, {Rix}, {Valentino}, {Wang}, \& {Williams}}]{deGraaff2024}
{de Graaff}, A., {Setton}, D.~J., {Brammer}, G., {et~al.} 2024, arXiv e-prints, arXiv:2404.05683

\bibitem[{{Draine}(2003)}]{Draine2003}
{Draine}, B.~T. 2003, \araa, 41, 241

\bibitem[{{Enia} {et~al.}(2022){Enia}, {Talia}, {Pozzi}, {Cimatti}, {Delvecchio}, {Zamorani}, {D'Amato}, {Bisigello}, {Gruppioni}, {Rodighiero}, {Calura}, {Dallacasa}, {Giulietti}, {Barchiesi}, {Behiri}, \& {Romano}}]{Enia2022}
{Enia}, A., {Talia}, M., {Pozzi}, F., {et~al.} 2022, \apj, 927, 204

\bibitem[{{Euclid Collaboration: Bisigello} {et~al.}(2023){Euclid Collaboration: Bisigello}, {Conselice}, {Baes}, {Bolzonella}, {Brescia}, {Cavuoti}, {Cucciati}, {Humphrey}, {Hunt}, {Maraston}, {Pozzetti}, {Tortora}, {van Mierlo}, {Aghanim}, {Auricchio}, {Baldi}, {Bender}, {Bodendorf}, {Bonino}, {Branchini}, {Brinchmann}, {Camera}, {Capobianco}, {Carbone}, {Carretero}, {Castander}, {Castellano}, {Cimatti}, {Congedo}, {Conversi}, {Copin}, {Corcione}, {Courbin}, {Cropper}, {Da Silva}, {Degaudenzi}, {Douspis}, {Dubath}, {Duncan}, {Dupac}, {Dusini}, {Farrens}, {Ferriol}, {Frailis}, {Franceschi}, {Franzetti}, {Fumana}, {Garilli}, {Gillard}, {Gillis}, {Giocoli}, {Grazian}, {Grupp}, {Guzzo}, {Haugan}, {Holmes}, {Hormuth}, {Hornstrup}, {Jahnke}, {K{\"u}mmel}, {Kermiche}, {Kiessling}, {Kilbinger}, {Kohley}, {Kunz}, {Kurki-Suonio}, {Ligori}, {Lilje}, {Lloro}, {Maiorano}, {Mansutti}, {Marggraf}, {Markovic}, {Marulli}, {Massey}, {Maurogordato}, {Medinaceli}, {Meneghetti}, {Merlin}, {Meylan}, {Moresco}, {Moscardini},
  {Munari}, {Niemi}, {Padilla}, {Paltani}, {Pasian}, {Pedersen}, {Pettorino}, {Polenta}, {Poncet}, {Popa}, {Raison}, {Renzi}, {Rhodes}, {Riccio}, {Rix}, {Romelli}, {Roncarelli}, {Rosset}, {Rossetti}, {Saglia}, {Sapone}, {Sartoris}, {Schneider}, {Scodeggio}, {Secroun}, {Seidel}, {Sirignano}, {Sirri}, {Stanco}, {Tallada-Cresp{\'\i}}, {Tavagnacco}, {Taylor}, {Tereno}, {Toledo-Moreo}, {Torradeflot}, {Tutusaus}, {Valentijn}, {Valenziano}, {Vassallo}, {Wang}, {Zacchei}, {Zamorani}, {Zoubian}, {Andreon}, {Bardelli}, {Boucaud}, {Colodro-Conde}, {Di Ferdinando}, {Graci{\'a}-Carpio}, {Lindholm}, {Maino}, {Mei}, {Scottez}, {Sureau}, {Tenti}, {Zucca}, {Borlaff}, {Ballardini}, {Biviano}, {Bozzo}, {Burigana}, {Cabanac}, {Cappi}, {Carvalho}, {Casas}, {Castignani}, {Cooray}, {Coupon}, {Courtois}, {Cuby}, {Davini}, {De Lucia}, {Desprez}, {Dole}, {Escartin}, {Escoffier}, {Farina}, {Fotopoulou}, {Ganga}, {Garcia-Bellido}, {George}, {Giacomini}, {Gozaliasl}, {Hildebrandt}, {Hook}, {Huertas-Company}, {Kansal}, {Keihanen},
  {Kirkpatrick}, {Loureiro}, {Mac{\'\i}as-P{\'e}rez}, {Magliocchetti}, {Mainetti}, {Marcin}, {Martinelli}, {Martinet}, {Metcalf}, {Monaco}, {Morgante}, {Nadathur}, {Nucita}, {Patrizii}, {Peel}, {Potter}, {Pourtsidou}, {P{\"o}ntinen}, {Reimberg}, {S{\'a}nchez}, {Sakr}, {Schirmer}, {Sefusatti}, {Sereno}, {Stadel}, {Teyssier}, {Valieri}, {Valiviita}, \& {Viel}}]{Bisigello2023}
{Euclid Collaboration: Bisigello}, L., {Conselice}, C.~J., {Baes}, M., {et~al.} 2023, \mnras, 520, 3529

\bibitem[{{Ferruit} {et~al.}(2022){Ferruit}, {Jakobsen}, {Giardino}, {Rawle}, {Alves de Oliveira}, {Arribas}, {Beck}, {Birkmann}, {B{\"o}ker}, {Bunker}, {Charlot}, {de Marchi}, {Franx}, {Henry}, {Karakla}, {Kassin}, {Kumari}, {L{\'o}pez-Caniego}, {L{\"u}tzgendorf}, {Maiolino}, {Manjavacas}, {Marston}, {Moseley}, {Muzerolle}, {Pirzkal}, {Rauscher}, {Rix}, {Sabbi}, {Sirianni}, {te Plate}, {Valenti}, {Willott}, \& {Zeidler}}]{Ferruit2022}
{Ferruit}, P., {Jakobsen}, P., {Giardino}, G., {et~al.} 2022, \aap, 661, A81

\bibitem[{{Finkelstein} {et~al.}(2023){Finkelstein}, {Bagley}, {Ferguson}, {Wilkins}, {Kartaltepe}, {Papovich}, {Yung}, {Arrabal Haro}, {Behroozi}, {Dickinson}, {Kocevski}, {Koekemoer}, {Larson}, {Le Bail}, {Morales}, {P{\'e}rez-Gonz{\'a}lez}, {Burgarella}, {Dav{\'e}}, {Hirschmann}, {Somerville}, {Wuyts}, {Bromm}, {Casey}, {Fontana}, {Fujimoto}, {Gardner}, {Giavalisco}, {Grazian}, {Grogin}, {Hathi}, {Hutchison}, {Jha}, {Jogee}, {Kewley}, {Kirkpatrick}, {Long}, {Lotz}, {Pentericci}, {Pierel}, {Pirzkal}, {Ravindranath}, {Ryan}, {Trump}, {Yang}, {Bhatawdekar}, {Bisigello}, {Buat}, {Calabr{\`o}}, {Castellano}, {Cleri}, {Cooper}, {Croton}, {Daddi}, {Dekel}, {Elbaz}, {Franco}, {Gawiser}, {Holwerda}, {Huertas-Company}, {Jaskot}, {Leung}, {Lucas}, {Mobasher}, {Pandya}, {Tacchella}, {Weiner}, \& {Zavala}}]{Finkelstein2023a}
{Finkelstein}, S.~L., {Bagley}, M.~B., {Ferguson}, H.~C., {et~al.} 2023, \apjl, 946, L13

\bibitem[{{Finkelstein} {et~al.}(2024){Finkelstein}, {Leung}, {Bagley}, {Dickinson}, {Ferguson}, {Papovich}, {Akins}, {Arrabal Haro}, {Dav{\'e}}, {Dekel}, {Kartaltepe}, {Kocevski}, {Koekemoer}, {Pirzkal}, {Somerville}, {Yung}, {Amor{\'\i}n}, {Backhaus}, {Behroozi}, {Bisigello}, {Bromm}, {Casey}, {Ch{\'a}vez Ortiz}, {Cheng}, {Chworowsky}, {Cleri}, {Cooper}, {Davis}, {de la Vega}, {Elbaz}, {Franco}, {Fontana}, {Fujimoto}, {Giavalisco}, {Grogin}, {Holwerda}, {Huertas-Company}, {Hirschmann}, {Iyer}, {Jogee}, {Jung}, {Larson}, {Lucas}, {Mobasher}, {Morales}, {Morley}, {Mukherjee}, {P{\'e}rez-Gonz{\'a}lez}, {Ravindranath}, {Rodighiero}, {Rowland}, {Tacchella}, {Taylor}, {Trump}, \& {Wilkins}}]{Finkelstein2024}
{Finkelstein}, S.~L., {Leung}, G. C.~K., {Bagley}, M.~B., {et~al.} 2024, \apjl, 969, L2

\bibitem[{{Franco} {et~al.}(2018){Franco}, {Elbaz}, {B{\'e}thermin}, {Magnelli}, {Schreiber}, {Ciesla}, {Dickinson}, {Nagar}, {Silverman}, {Daddi}, {Alexander}, {Wang}, {Pannella}, {Le Floc'h}, {Pope}, {Giavalisco}, {Maury}, {Bournaud}, {Chary}, {Demarco}, {Ferguson}, {Finkelstein}, {Inami}, {Iono}, {Juneau}, {Lagache}, {Leiton}, {Lin}, {Magdis}, {Messias}, {Motohara}, {Mullaney}, {Okumura}, {Papovich}, {Pforr}, {Rujopakarn}, {Sargent}, {Shu}, \& {Zhou}}]{Franco2018}
{Franco}, M., {Elbaz}, D., {B{\'e}thermin}, M., {et~al.} 2018, \aap, 620, A152

\bibitem[{{Furtak} {et~al.}(2023){Furtak}, {Zitrin}, {Plat}, {Fujimoto}, {Wang}, {Nelson}, {Labb{\'e}}, {Bezanson}, {Brammer}, {van Dokkum}, {Endsley}, {Glazebrook}, {Greene}, {Leja}, {Price}, {Smit}, {Stark}, {Weaver}, {Whitaker}, {Atek}, {Chevallard}, {Curtis-Lake}, {Dayal}, {Feltre}, {Franx}, {Fudamoto}, {Marchesini}, {Mowla}, {Pan}, {Suess}, {Vidal-Garc{\'\i}a}, \& {Williams}}]{Furtak2023}
{Furtak}, L.~J., {Zitrin}, A., {Plat}, A., {et~al.} 2023, \apj, 952, 142

\bibitem[{{Gall} {et~al.}(2011){Gall}, {Hjorth}, \& {Andersen}}]{Gall2011}
{Gall}, C., {Hjorth}, J., \& {Andersen}, A.~C. 2011, \aapr, 19, 43

\bibitem[{{Gentile} {et~al.}(2024{\natexlab{a}}){Gentile}, {Talia}, {Behiri}, {Zamorani}, {Barchiesi}, {Vignali}, {Pozzi}, {Bethermin}, {Enia}, {Faisst}, {Giulietti}, {Gruppioni}, {Lapi}, {Massardi}, {Smol{\v{c}}i{\'c}}, {Vaccari}, \& {Cimatti}}]{Gentile2024}
{Gentile}, F., {Talia}, M., {Behiri}, M., {et~al.} 2024{\natexlab{a}}, \apj, 962, 26

\bibitem[{{Gentile} {et~al.}(2024{\natexlab{b}}){Gentile}, {Talia}, {Daddi}, {Giulietti}, {Lapi}, {Massardi}, {Pozzi}, {Zamorani}, {Behiri}, {Enia}, {Bethermin}, {Dallacasa}, {Delvecchio}, {Faisst}, {Gruppioni}, {Loiacono}, {Traina}, {Vaccari}, {Vallini}, {Vignali}, {Smol{\v{c}}i{\'c}}, \& {Cimatti}}]{Gentile2024b}
{Gentile}, F., {Talia}, M., {Daddi}, E., {et~al.} 2024{\natexlab{b}}, \aap, 687, A288

\bibitem[{{Gonz{\'a}lez Delgado} {et~al.}(2005){Gonz{\'a}lez Delgado}, {Cervi{\~n}o}, {Martins}, {Leitherer}, \& {Hauschildt}}]{GonzalezDelgado2005}
{Gonz{\'a}lez Delgado}, R.~M., {Cervi{\~n}o}, M., {Martins}, L.~P., {Leitherer}, C., \& {Hauschildt}, P.~H. 2005, \mnras, 357, 945

\bibitem[{Grogin {et~al.}(2011)Grogin, Kocevski, Faber, Ferguson, Koekemoer, Riess, Acquaviva, Alexander, Almaini, Ashby, Barden, Bell, Bournaud, Brown, Caputi, Casertano, Cassata, Castellano, Challis, Chary, Cheung, Cirasuolo, Conselice, Cooray, Croton, Daddi, Dahlen, Dav{\'{e}}, de~Mello, Dekel, Dickinson, Dolch, Donley, Dunlop, Dutton, Elbaz, Fazio, Filippenko, Finkelstein, Fontana, Gardner, Garnavich, Gawiser, Giavalisco, Grazian, Guo, Hathi, Häussler, Hopkins, Huang, Huang, Jha, Kartaltepe, Kirshner, Koo, Lai, Lee, Li, Lotz, Lucas, Madau, McCarthy, McGrath, McIntosh, McLure, Mobasher, Moustakas, Mozena, Nandra, Newman, Niemi, Noeske, Papovich, Pentericci, Pope, Primack, Rajan, Ravindranath, Reddy, Renzini, Rix, Robaina, Rodney, Rosario, Rosati, Salimbeni, Scarlata, Siana, Simard, Smidt, Somerville, Spinrad, Straughn, Strolger, Telford, Teplitz, Trump, van~der Wel, Villforth, Wechsler, Weiner, Wiklind, Wild, Wilson, Wuyts, Yan, \& Yun}]{Grogin2011}
Grogin, N.~A., Kocevski, D.~D., Faber, S.~M., {et~al.} 2011, The Astrophysical Journal Supplement Series, 197, 35

\bibitem[{{Gruppioni} {et~al.}(2020){Gruppioni}, {B{\'e}thermin}, {Loiacono}, {Le F{\`e}vre}, {Capak}, {Cassata}, {Faisst}, {Schaerer}, {Silverman}, {Yan}, {Bardelli}, {Boquien}, {Carraro}, {Cimatti}, {Dessauges-Zavadsky}, {Ginolfi}, {Fujimoto}, {Hathi}, {Jones}, {Khusanova}, {Koekemoer}, {Lagache}, {Lemaux}, {Oesch}, {Pozzi}, {Riechers}, {Rodighiero}, {Romano}, {Talia}, {Vallini}, {Vergani}, {Zamorani}, \& {Zucca}}]{Gruppioni2020}
{Gruppioni}, C., {B{\'e}thermin}, M., {Loiacono}, F., {et~al.} 2020, \aap, 643, A8

\bibitem[{{Gruppioni} {et~al.}(2013){Gruppioni}, {Pozzi}, {Rodighiero}, {Delvecchio}, {Berta}, {Pozzetti}, {Zamorani}, {Andreani}, {Cimatti}, {Ilbert}, {Le Floc'h}, {Lutz}, {Magnelli}, {Marchetti}, {Monaco}, {Nordon}, {Oliver}, {Popesso}, {Riguccini}, {Roseboom}, {Rosario}, {Sargent}, {Vaccari}, {Altieri}, {Aussel}, {Bongiovanni}, {Cepa}, {Daddi}, {Dom{\'\i}nguez-S{\'a}nchez}, {Elbaz}, {F{\"o}rster Schreiber}, {Genzel}, {Iribarrem}, {Magliocchetti}, {Maiolino}, {Poglitsch}, {P{\'e}rez Garc{\'\i}a}, {Sanchez-Portal}, {Sturm}, {Tacconi}, {Valtchanov}, {Amblard}, {Arumugam}, {Bethermin}, {Bock}, {Boselli}, {Buat}, {Burgarella}, {Castro-Rodr{\'\i}guez}, {Cava}, {Chanial}, {Clements}, {Conley}, {Cooray}, {Dowell}, {Dwek}, {Eales}, {Franceschini}, {Glenn}, {Griffin}, {Hatziminaoglou}, {Ibar}, {Isaak}, {Ivison}, {Lagache}, {Levenson}, {Lu}, {Madden}, {Maffei}, {Mainetti}, {Nguyen}, {O'Halloran}, {Page}, {Panuzzo}, {Papageorgiou}, {Pearson}, {P{\'e}rez-Fournon}, {Pohlen}, {Rigopoulou}, {Rowan-Robinson}, {Schulz},
  {Scott}, {Seymour}, {Shupe}, {Smith}, {Stevens}, {Symeonidis}, {Trichas}, {Tugwell}, {Vigroux}, {Wang}, {Wright}, {Xu}, {Zemcov}, {Bardelli}, {Carollo}, {Contini}, {Le F{\'e}vre}, {Lilly}, {Mainieri}, {Renzini}, {Scodeggio}, \& {Zucca}}]{Gruppioni2013}
{Gruppioni}, C., {Pozzi}, F., {Rodighiero}, G., {et~al.} 2013, \mnras, 432, 23

\bibitem[{{Guia} {et~al.}(2024){Guia}, {Pacucci}, \& {Kocevski}}]{Guia2024}
{Guia}, C.~A., {Pacucci}, F., \& {Kocevski}, D.~D. 2024, Research Notes of the American Astronomical Society, 8, 207

\bibitem[{{Horne}(1986)}]{Horne1986}
{Horne}, K. 1986, \pasp, 98, 609

\bibitem[{{JDADF Developers} {et~al.}(2023){JDADF Developers}, {Averbukh}, {Bradley}, {Buikhuizen}, \& et~al.}]{JDADF2023}
{JDADF Developers}, {Averbukh}, J., {Bradley}, L., {Buikhuizen}, M., \& et~al. 2023

\bibitem[{{Kauffmann} {et~al.}(2003){Kauffmann}, {Heckman}, {Tremonti}, {Brinchmann}, {Charlot}, {White}, {Ridgway}, {Brinkmann}, {Fukugita}, {Hall}, {Ivezi{\'c}}, {Richards}, \& {Schneider}}]{Kauffmann2003}
{Kauffmann}, G., {Heckman}, T.~M., {Tremonti}, C., {et~al.} 2003, \mnras, 346, 1055

\bibitem[{{Kennicutt} \& {Evans}(2012)}]{Kennicutt2012}
{Kennicutt}, R.~C. \& {Evans}, N.~J. 2012, \araa, 50, 531

\bibitem[{{Kewley} {et~al.}(2013){Kewley}, {Dopita}, {Leitherer}, {Dav{\'e}}, {Yuan}, {Allen}, {Groves}, \& {Sutherland}}]{Kewley2013}
{Kewley}, L.~J., {Dopita}, M.~A., {Leitherer}, C., {et~al.} 2013, \apj, 774, 100

\bibitem[{{Kewley} {et~al.}(2006){Kewley}, {Groves}, {Kauffmann}, \& {Heckman}}]{Kewley2006}
{Kewley}, L.~J., {Groves}, B., {Kauffmann}, G., \& {Heckman}, T. 2006, \mnras, 372, 961

\bibitem[{{Koekemoer} {et~al.}(2011){Koekemoer}, {Faber}, {Ferguson}, {Grogin}, {Kocevski}, {Koo}, {Lai}, {Lotz}, {Lucas}, {McGrath}, {Ogaz}, {Rajan}, {Riess}, {Rodney}, {Strolger}, {Casertano}, {Castellano}, {Dahlen}, {Dickinson}, {Dolch}, {Fontana}, {Giavalisco}, {Grazian}, {Guo}, {Hathi}, {Huang}, {van der Wel}, {Yan}, {Acquaviva}, {Alexander}, {Almaini}, {Ashby}, {Barden}, {Bell}, {Bournaud}, {Brown}, {Caputi}, {Cassata}, {Challis}, {Chary}, {Cheung}, {Cirasuolo}, {Conselice}, {Roshan Cooray}, {Croton}, {Daddi}, {Dav{\'e}}, {de Mello}, {de Ravel}, {Dekel}, {Donley}, {Dunlop}, {Dutton}, {Elbaz}, {Fazio}, {Filippenko}, {Finkelstein}, {Frazer}, {Gardner}, {Garnavich}, {Gawiser}, {Gruetzbauch}, {Hartley}, {H{\"a}ussler}, {Herrington}, {Hopkins}, {Huang}, {Jha}, {Johnson}, {Kartaltepe}, {Khostovan}, {Kirshner}, {Lani}, {Lee}, {Li}, {Madau}, {McCarthy}, {McIntosh}, {McLure}, {McPartland}, {Mobasher}, {Moreira}, {Mortlock}, {Moustakas}, {Mozena}, {Nandra}, {Newman}, {Nielsen}, {Niemi}, {Noeske}, {Papovich},
  {Pentericci}, {Pope}, {Primack}, {Ravindranath}, {Reddy}, {Renzini}, {Rix}, {Robaina}, {Rosario}, {Rosati}, {Salimbeni}, {Scarlata}, {Siana}, {Simard}, {Smidt}, {Snyder}, {Somerville}, {Spinrad}, {Straughn}, {Telford}, {Teplitz}, {Trump}, {Vargas}, {Villforth}, {Wagner}, {Wandro}, {Wechsler}, {Weiner}, {Wiklind}, {Wild}, {Wilson}, {Wuyts}, \& {Yun}}]{Koekemoer2011}
{Koekemoer}, A.~M., {Faber}, S.~M., {Ferguson}, H.~C., {et~al.} 2011, \apjs, 197, 36

\bibitem[{{Kreckel} {et~al.}(2013){Kreckel}, {Groves}, {Schinnerer}, {Johnson}, {Aniano}, {Calzetti}, {Croxall}, {Draine}, {Gordon}, {Crocker}, {Dale}, {Hunt}, {Kennicutt}, {Meidt}, {Smith}, \& {Tabatabaei}}]{Kreckel2013}
{Kreckel}, K., {Groves}, B., {Schinnerer}, E., {et~al.} 2013, \apj, 771, 62

\bibitem[{{Kroupa}(2001)}]{Kroupa2001}
{Kroupa}, P. 2001, \mnras, 322, 231

\bibitem[{{Leja} {et~al.}(2019){Leja}, {Carnall}, {Johnson}, {Conroy}, \& {Speagle}}]{Leja2019}
{Leja}, J., {Carnall}, A.~C., {Johnson}, B.~D., {Conroy}, C., \& {Speagle}, J.~S. 2019, \apj, 876, 3

\bibitem[{{Liu} {et~al.}(2024){Liu}, {Morishita}, \& {Kodama}}]{Liu2024}
{Liu}, Z., {Morishita}, T., \& {Kodama}, T. 2024, arXiv e-prints, arXiv:2406.11188

\bibitem[{{Lo Faro} {et~al.}(2017){Lo Faro}, {Buat}, {Roehlly}, {Alvarez-Marquez}, {Burgarella}, {Silva}, \& {Efstathiou}}]{Lofaro2017}
{Lo Faro}, B., {Buat}, V., {Roehlly}, Y., {et~al.} 2017, \mnras, 472, 1372

\bibitem[{{Lorenz} {et~al.}(2024){Lorenz}, {Kriek}, {Shapley}, {Sanders}, {Coil}, {Leja}, {Mobasher}, {Nelson}, {Price}, {Reddy}, {Runco}, {Suess}, {Shivaei}, {Siana}, \& {Weisz}}]{Lorenz2024}
{Lorenz}, B., {Kriek}, M., {Shapley}, A.~E., {et~al.} 2024, \apj, 975, 187

\bibitem[{{Magnelli} {et~al.}(2024){Magnelli}, {Adscheid}, {Wang}, {Ciesla}, {Daddi}, {Delvecchio}, {Elbaz}, {Fudamoto}, {Fukushima}, {Franco}, {G{\'o}mez-Guijarro}, {Gruppioni}, {Jim{\'e}nez-Andrade}, {Liu}, {Oesch}, {Schinnerer}, \& {Traina}}]{Magnelli2024}
{Magnelli}, B., {Adscheid}, S., {Wang}, T.-M., {et~al.} 2024, \aap, 688, A55

\bibitem[{{Maiolino} \& {Mannucci}(2019)}]{Maiolino2019}
{Maiolino}, R. \& {Mannucci}, F. 2019, \aapr, 27, 3

\bibitem[{{Maiolino} {et~al.}(2024){Maiolino}, {Scholtz}, {Curtis-Lake}, {Carniani}, {Baker}, {de Graaff}, {Tacchella}, {{\"U}bler}, {D'Eugenio}, {Witstok}, {Curti}, {Arribas}, {Bunker}, {Charlot}, {Chevallard}, {Eisenstein}, {Egami}, {Ji}, {Jones}, {Lyu}, {Rawle}, {Robertson}, {Rujopakarn}, {Perna}, {Sun}, {Venturi}, {Williams}, \& {Willott}}]{Maiolino2023}
{Maiolino}, R., {Scholtz}, J., {Curtis-Lake}, E., {et~al.} 2024, \aap, 691, A145

\bibitem[{{Mannucci} {et~al.}(2010){Mannucci}, {Cresci}, {Maiolino}, {Marconi}, \& {Gnerucci}}]{Mannucci2010}
{Mannucci}, F., {Cresci}, G., {Maiolino}, R., {Marconi}, A., \& {Gnerucci}, A. 2010, \mnras, 408, 2115

\bibitem[{{Matthee} {et~al.}(2024){Matthee}, {Naidu}, {Brammer}, {Chisholm}, {Eilers}, {Goulding}, {Greene}, {Kashino}, {Labbe}, {Lilly}, {Mackenzie}, {Oesch}, {Weibel}, {Wuyts}, {Xiao}, {Bordoloi}, {Bouwens}, {van Dokkum}, {Illingworth}, {Kramarenko}, {Maseda}, {Mason}, {Meyer}, {Nelson}, {Reddy}, {Shivaei}, {Simcoe}, \& {Yue}}]{Matthee2024}
{Matthee}, J., {Naidu}, R.~P., {Brammer}, G., {et~al.} 2024, \apj, 963, 129

\bibitem[{{McLure} {et~al.}(2018){McLure}, {Dunlop}, {Cullen}, {Bourne}, {Best}, {Khochfar}, {Bowler}, {Biggs}, {Geach}, {Scott}, {Micha{\l}owski}, {Rujopakarn}, {van Kampen}, {Kirkpatrick}, \& {Pope}}]{McLure2018}
{McLure}, R.~J., {Dunlop}, J.~S., {Cullen}, F., {et~al.} 2018, \mnras, 476, 3991

\bibitem[{{Naidu} {et~al.}(2022){Naidu}, {Oesch}, {Setton}, {Matthee}, {Conroy}, {Johnson}, {Weaver}, {Bouwens}, {Brammer}, {Dayal}, {Illingworth}, {Barrufet}, {Belli}, {Bezanson}, {Bose}, {Heintz}, {Leja}, {Leonova}, {Marques-Chaves}, {Stefanon}, {Toft}, {van der Wel}, {van Dokkum}, {Weibel}, \& {Whitaker}}]{Naidu2022}
{Naidu}, R.~P., {Oesch}, P.~A., {Setton}, D.~J., {et~al.} 2022, arXiv e-prints, arXiv:2208.02794

\bibitem[{{Nakajima} {et~al.}(2022){Nakajima}, {Ouchi}, {Xu}, {Rauch}, {Harikane}, {Nishigaki}, {Isobe}, {Kusakabe}, {Nagao}, {Ono}, {Onodera}, {Sugahara}, {Kim}, {Komiyama}, {Lee}, \& {Zahedy}}]{Nakajima2022}
{Nakajima}, K., {Ouchi}, M., {Xu}, Y., {et~al.} 2022, \apjs, 262, 3

\bibitem[{{Oke} \& {Gunn}(1983)}]{Oke1983}
{Oke}, J.~B. \& {Gunn}, J.~E. 1983, \apj, 266, 713

\bibitem[{{Ono} {et~al.}(2024){Ono}, {Harikane}, {Ouchi}, {Nakajima}, {Isobe}, {Shibuya}, {Nakane}, {Umeda}, {Xu}, \& {Zhang}}]{Ono2024}
{Ono}, Y., {Harikane}, Y., {Ouchi}, M., {et~al.} 2024, \pasj, 76, 219

\bibitem[{{Pannella} {et~al.}(2015){Pannella}, {Elbaz}, {Daddi}, {Dickinson}, {Hwang}, {Schreiber}, {Strazzullo}, {Aussel}, {Bethermin}, {Buat}, {Charmandaris}, {Cibinel}, {Juneau}, {Ivison}, {Le Borgne}, {Le Floc'h}, {Leiton}, {Lin}, {Magdis}, {Morrison}, {Mullaney}, {Onodera}, {Renzini}, {Salim}, {Sargent}, {Scott}, {Shu}, \& {Wang}}]{Pannella2015}
{Pannella}, M., {Elbaz}, D., {Daddi}, E., {et~al.} 2015, \apj, 807, 141

\bibitem[{{P{\'e}rez-Gonz{\'a}lez} {et~al.}(2023){P{\'e}rez-Gonz{\'a}lez}, {Barro}, {Annunziatella}, {Costantin}, {Garc{\'\i}a-Argum{\'a}nez}, {McGrath}, {M{\'e}rida}, {Zavala}, {Arrabal Haro}, {Bagley}, {Backhaus}, {Behroozi}, {Bell}, {Bisigello}, {Buat}, {Calabr{\`o}}, {Casey}, {Cleri}, {Coogan}, {Cooper}, {Cooray}, {Dekel}, {Dickinson}, {Elbaz}, {Ferguson}, {Finkelstein}, {Fontana}, {Franco}, {Gardner}, {Giavalisco}, {G{\'o}mez-Guijarro}, {Grazian}, {Grogin}, {Guo}, {Huertas-Company}, {Jogee}, {Kartaltepe}, {Kewley}, {Kirkpatrick}, {Kocevski}, {Koekemoer}, {Long}, {Lotz}, {Lucas}, {Papovich}, {Pirzkal}, {Ravindranath}, {Somerville}, {Tacchella}, {Trump}, {Wang}, {Wilkins}, {Wuyts}, {Yang}, \& {Yung}}]{PerezGonzalez2023}
{P{\'e}rez-Gonz{\'a}lez}, P.~G., {Barro}, G., {Annunziatella}, M., {et~al.} 2023, \apjl, 946, L16

\bibitem[{{P{\'e}rez-Montero}(2017)}]{PerezMontero2017}
{P{\'e}rez-Montero}, E. 2017, \pasp, 129, 043001

\bibitem[{{Perrin} {et~al.}(2012){Perrin}, {Soummer}, {Elliott}, {Lallo}, \& {Sivaramakrishnan}}]{Perrin2012}
{Perrin}, M.~D., {Soummer}, R., {Elliott}, E.~M., {Lallo}, M.~D., \& {Sivaramakrishnan}, A. 2012, in Society of Photo-Optical Instrumentation Engineers (SPIE) Conference Series, Vol. 8442, Space Telescopes and Instrumentation 2012: Optical, Infrared, and Millimeter Wave, ed. M.~C. {Clampin}, G.~G. {Fazio}, H.~A. {MacEwen}, \& J.~{Oschmann}, Jacobus~M., 84423D

\bibitem[{Pettini \& Pagel(2004)}]{Pettini2004}
Pettini, M. \& Pagel, B. E.~J. 2004, \mnras, 348, L59

\bibitem[{{Pilyugin} {et~al.}(2009){Pilyugin}, {Mattsson}, {V{\'\i}lchez}, \& {Cedr{\'e}s}}]{Pilyugin2009}
{Pilyugin}, L.~S., {Mattsson}, L., {V{\'\i}lchez}, J.~M., \& {Cedr{\'e}s}, B. 2009, \mnras, 398, 485

\bibitem[{{Puglisi} {et~al.}(2019){Puglisi}, {Daddi}, {Liu}, {Bournaud}, {Silverman}, {Circosta}, {Calabr{\`o}}, {Aravena}, {Cibinel}, {Dannerbauer}, {Delvecchio}, {Elbaz}, {Gao}, {Gobat}, {Jin}, {Le Floc'h}, {Magdis}, {Mancini}, {Riechers}, {Rodighiero}, {Sargent}, {Valentino}, \& {Zanisi}}]{Puglisi2019}
{Puglisi}, A., {Daddi}, E., {Liu}, D., {et~al.} 2019, \apjl, 877, L23

\bibitem[{{Rodighiero} {et~al.}(2023){Rodighiero}, {Bisigello}, {Iani}, {Marasco}, {Grazian}, {Sinigaglia}, {Cassata}, \& {Gruppioni}}]{Rodighiero2023}
{Rodighiero}, G., {Bisigello}, L., {Iani}, E., {et~al.} 2023, \mnras, 518, L19

\bibitem[{{Rodighiero} {et~al.}(2007){Rodighiero}, {Cimatti}, {Franceschini}, {Brusa}, {Fritz}, \& {Bolzonella}}]{Rodighiero2007}
{Rodighiero}, G., {Cimatti}, A., {Franceschini}, A., {et~al.} 2007, \aap, 470, 21

\bibitem[{{Rodr{\'\i}guez-Mu{\~n}oz} {et~al.}(2022){Rodr{\'\i}guez-Mu{\~n}oz}, {Rodighiero}, {P{\'e}rez-Gonz{\'a}lez}, {Talia}, {Baronchelli}, {Morselli}, {Renzini}, {Puglisi}, {Grazian}, {Zanella}, {Mancini}, {Feltre}, {Romano}, {Vidal Garc{\'\i}a}, {Franceschini}, {Alcalde Pampliega}, {Cassata}, {Costantin}, {Dom{\'\i}nguez S{\'a}nchez}, {Espino-Briones}, {Iani}, {Koekemoer}, {Lumbreras-Calle}, \& {Rodr{\'\i}guez-Espinosa}}]{RodriguezMunoz2022}
{Rodr{\'\i}guez-Mu{\~n}oz}, L., {Rodighiero}, G., {P{\'e}rez-Gonz{\'a}lez}, P.~G., {et~al.} 2022, \mnras, 510, 2061

\bibitem[{{Roper} {et~al.}(2023){Roper}, {Lovell}, {Vijayan}, {Irodotou}, {Kuusisto}, {Matharu}, {Seeyave}, {Thomas}, \& {Wilkins}}]{Roper2023}
{Roper}, W.~J., {Lovell}, C.~C., {Vijayan}, A.~P., {et~al.} 2023, \mnras, 526, 6128

\bibitem[{{Salmon} {et~al.}(2016){Salmon}, {Papovich}, {Long}, {Willner}, {Finkelstein}, {Ferguson}, {Dickinson}, {Duncan}, {Faber}, {Hathi}, {Koekemoer}, {Kurczynski}, {Newman}, {Pacifici}, {P{\'e}rez-Gonz{\'a}lez}, \& {Pforr}}]{Salmon2016}
{Salmon}, B., {Papovich}, C., {Long}, J., {et~al.} 2016, \apj, 827, 20

\bibitem[{{Santini} {et~al.}(2017){Santini}, {Fontana}, {Castellano}, {Di Criscienzo}, {Merlin}, {Amorin}, {Cullen}, {Daddi}, {Dickinson}, {Dunlop}, {Grazian}, {Lamastra}, {McLure}, {Micha{\l}owski}, {Pentericci}, \& {Shu}}]{Santini2017}
{Santini}, P., {Fontana}, A., {Castellano}, M., {et~al.} 2017, \apj, 847, 76

\bibitem[{{Sarangi} {et~al.}(2018){Sarangi}, {Matsuura}, \& {Micelotta}}]{Sarangi2018}
{Sarangi}, A., {Matsuura}, M., \& {Micelotta}, E.~R. 2018, \ssr, 214, 63

\bibitem[{{Seill{\'e}} {et~al.}(2022){Seill{\'e}}, {Buat}, {Haddad}, {Boselli}, {Boquien}, {Ciesla}, {Roehlly}, \& {Burgarella}}]{Seille2022}
{Seill{\'e}}, L.~M., {Buat}, V., {Haddad}, W., {et~al.} 2022, \aap, 665, A137

\bibitem[{{Sersic}(1968)}]{Sersic1968}
{Sersic}, J.~L. 1968, {Atlas de Galaxias Australes}

\bibitem[{{Shapley} {et~al.}(2023){Shapley}, {Sanders}, {Reddy}, {Topping}, \& {Brammer}}]{Shapley2023}
{Shapley}, A.~E., {Sanders}, R.~L., {Reddy}, N.~A., {Topping}, M.~W., \& {Brammer}, G.~B. 2023, \apj, 954, 157

\bibitem[{{Shapley} {et~al.}(2022){Shapley}, {Sanders}, {Salim}, {Reddy}, {Kriek}, {Mobasher}, {Coil}, {Siana}, {Price}, {Shivaei}, {Dunlop}, {McLure}, \& {Cullen}}]{Shapley2022}
{Shapley}, A.~E., {Sanders}, R.~L., {Salim}, S., {et~al.} 2022, \apj, 926, 145

\bibitem[{{Shirazi} \& {Brinchmann}(2012)}]{Shirazi2012}
{Shirazi}, M. \& {Brinchmann}, J. 2012, \mnras, 421, 1043

\bibitem[{{Simpson} {et~al.}(2014){Simpson}, {Swinbank}, {Smail}, {Alexander}, {Brandt}, {Bertoldi}, {de Breuck}, {Chapman}, {Coppin}, {da Cunha}, {Danielson}, {Dannerbauer}, {Greve}, {Hodge}, {Ivison}, {Karim}, {Knudsen}, {Poggianti}, {Schinnerer}, {Thomson}, {Walter}, {Wardlow}, {Wei{\ss}}, \& {van der Werf}}]{Simpson2014}
{Simpson}, J.~M., {Swinbank}, A.~M., {Smail}, I., {et~al.} 2014, \apj, 788, 125

\bibitem[{{Tacchella} {et~al.}(2022){Tacchella}, {Finkelstein}, {Bagley}, {Dickinson}, {Ferguson}, {Giavalisco}, {Graziani}, {Grogin}, {Hathi}, {Hutchison}, {Jung}, {Koekemoer}, {Larson}, {Papovich}, {Pirzkal}, {Rojas-Ruiz}, {Song}, {Schneider}, {Somerville}, {Wilkins}, \& {Yung}}]{Tacchella2022}
{Tacchella}, S., {Finkelstein}, S.~L., {Bagley}, M., {et~al.} 2022, \apj, 927, 170

\bibitem[{{Talia} {et~al.}(2021){Talia}, {Cimatti}, {Giulietti}, {Zamorani}, {Bethermin}, {Faisst}, {Le F{\`e}vre}, \& {Smol{\c{c}}i{\'c}}}]{Talia2021}
{Talia}, M., {Cimatti}, A., {Giulietti}, M., {et~al.} 2021, \apj, 909, 23

\bibitem[{{Talia} {et~al.}(2015){Talia}, {Cimatti}, {Pozzetti}, {Rodighiero}, {Gruppioni}, {Pozzi}, {Daddi}, {Maraston}, {Mignoli}, \& {Kurk}}]{Talia2015}
{Talia}, M., {Cimatti}, A., {Pozzetti}, L., {et~al.} 2015, \aap, 582, A80

\bibitem[{{Traina} {et~al.}(2024{\natexlab{a}}){Traina}, {Gruppioni}, {Delvecchio}, {Calura}, {Bisigello}, {Feltre}, {Magnelli}, {Schinnerer}, {Liu}, {Adscheid}, {Behiri}, {Gentile}, {Pozzi}, {Talia}, {Zamorani}, {Algera}, {Gillman}, {Lambrides}, \& {Symeonidis}}]{Traina2024}
{Traina}, A., {Gruppioni}, C., {Delvecchio}, I., {et~al.} 2024{\natexlab{a}}, \aap, 681, A118

\bibitem[{{Traina} {et~al.}(2024{\natexlab{b}}){Traina}, {Magnelli}, {Gruppioni}, {Delvecchio}, {Parente}, {Calura}, {Bisigello}, {Feltre}, {Pozzi}, \& {Vallini}}]{Traina2024b}
{Traina}, A., {Magnelli}, B., {Gruppioni}, C., {et~al.} 2024{\natexlab{b}}, \aap, 690, A84

\bibitem[{{Tremonti} {et~al.}(2004){Tremonti}, {Heckman}, {Kauffmann}, {Brinchmann}, {Charlot}, {White}, {Seibert}, {Peng}, {Schlegel}, {Uomoto}, {Fukugita}, \& {Brinkmann}}]{Tremonti2004}
{Tremonti}, C.~A., {Heckman}, T.~M., {Kauffmann}, G., {et~al.} 2004, \apj, 613, 898

\bibitem[{{{\"U}bler} {et~al.}(2023){{\"U}bler}, {Maiolino}, {Curtis-Lake}, {P{\'e}rez-Gonz{\'a}lez}, {Curti}, {Perna}, {Arribas}, {Charlot}, {Marshall}, {D'Eugenio}, {Scholtz}, {Bunker}, {Carniani}, {Ferruit}, {Jakobsen}, {Rix}, {Rodr{\'\i}guez Del Pino}, {Willott}, {Boeker}, {Cresci}, {Jones}, {Kumari}, \& {Rawle}}]{Ubler2023}
{{\"U}bler}, H., {Maiolino}, R., {Curtis-Lake}, E., {et~al.} 2023, \aap, 677, A145

\bibitem[{{Vijayan} {et~al.}(2021){Vijayan}, {Lovell}, {Wilkins}, {Thomas}, {Barnes}, {Irodotou}, {Kuusisto}, \& {Roper}}]{Vijayan2021}
{Vijayan}, A.~P., {Lovell}, C.~C., {Wilkins}, S.~M., {et~al.} 2021, \mnras, 501, 3289

\bibitem[{{Wang} {et~al.}(2016){Wang}, {Elbaz}, {Schreiber}, {Pannella}, {Shu}, {Willner}, {Ashby}, {Huang}, {Fontana}, {Dekel}, {Daddi}, {Ferguson}, {Dunlop}, {Ciesla}, {Koekemoer}, {Giavalisco}, {Boutsia}, {Finkelstein}, {Juneau}, {Barro}, {Koo}, {Micha{\l}owski}, {Orellana}, {Lu}, {Castellano}, {Bourne}, {Buitrago}, {Santini}, {Faber}, {Hathi}, {Lucas}, \& {P{\'e}rez-Gonz{\'a}lez}}]{WangT2016}
{Wang}, T., {Elbaz}, D., {Schreiber}, C., {et~al.} 2016, \apj, 816, 84

\bibitem[{{Wang} {et~al.}(2019){Wang}, {Schreiber}, {Elbaz}, {Yoshimura}, {Kohno}, {Shu}, {Yamaguchi}, {Pannella}, {Franco}, {Huang}, {Lim}, \& {Wang}}]{Wang2019}
{Wang}, T., {Schreiber}, C., {Elbaz}, D., {et~al.} 2019, \nat, 572, 211

\bibitem[{{Wang} {et~al.}(2024){Wang}, {Sun}, {Zhou}, {Xu}, {Cheng}, {Li}, {Chen}, {Mo}, {Dekel}, {Zheng}, {Cai}, {Yang}, {Dai}, {Elbaz}, \& {Huang}}]{Wang2024}
{Wang}, T., {Sun}, H., {Zhou}, L., {et~al.} 2024, arXiv e-prints, arXiv:2403.02399

\bibitem[{{Ward} {et~al.}(2024){Ward}, {de la Vega}, {Mobasher}, {McGrath}, {Iyer}, {Calabr{\`o}}, {Costantin}, {Dickinson}, {Holwerda}, {Huertas-Company}, {Hirschmann}, {Lucas}, {Pandya}, {Wilkins}, {Yung}, {Arrabal Haro}, {Bagley}, {Finkelstein}, {Kartaltepe}, {Koekemoer}, {Papovich}, \& {Pirzkal}}]{Ward2024}
{Ward}, E., {de la Vega}, A., {Mobasher}, B., {et~al.} 2024, \apj, 962, 176

\bibitem[{{Williams} {et~al.}(2024){Williams}, {Alberts}, {Ji}, {Hainline}, {Lyu}, {Rieke}, {Endsley}, {Suess}, {Sun}, {Johnson}, {Florian}, {Shivaei}, {Rujopakarn}, {Baker}, {Bhatawdekar}, {Boyett}, {Bunker}, {Cameron}, {Carniani}, {Charlot}, {Curtis-Lake}, {DeCoursey}, {de Graaff}, {Egami}, {Eisenstein}, {Gibson}, {Hausen}, {Helton}, {Maiolino}, {Maseda}, {Nelson}, {P{\'e}rez-Gonz{\'a}lez}, {Rieke}, {Robertson}, {Saxena}, {Tacchella}, {Willmer}, \& {Willott}}]{Williams2024}
{Williams}, C.~C., {Alberts}, S., {Ji}, Z., {et~al.} 2024, \apj, 968, 34

\bibitem[{{Xiao} {et~al.}(2023){Xiao}, {Oesch}, {Elbaz}, {Bing}, {Nelson}, {Weibel}, {Naidu}, {Daddi}, {Bouwens}, {Matthee}, {Wuyts}, {Chisholm}, {Brammer}, {Dickinson}, {Magnelli}, {Leroy}, {van Dokkum}, {Schaerer}, {Herard-Demanche}, {Barrufet}, {Endsley}, {Fudamoto}, {G{\'o}mez-Guijarro}, {Gottumukkala}, {Illingworth}, {Labbe}, {Magee}, {Marchesini}, {Maseda}, {Qin}, {Reddy}, {Shapley}, {Shivaei}, {Shuntov}, {Stefanon}, {Whitaker}, \& {Wyithe}}]{Xiao2023}
{Xiao}, M., {Oesch}, P., {Elbaz}, D., {et~al.} 2023, arXiv e-prints, arXiv:2309.02492

\bibitem[{{Zavala} {et~al.}(2017){Zavala}, {Aretxaga}, {Geach}, {Hughes}, {Birkinshaw}, {Chapin}, {Chapman}, {Chen}, {Clements}, {Dunlop}, {Farrah}, {Ivison}, {Jenness}, {Micha{\l}owski}, {Robson}, {Scott}, {Simpson}, {Spaans}, \& {van der Werf}}]{Zavala2017}
{Zavala}, J.~A., {Aretxaga}, I., {Geach}, J.~E., {et~al.} 2017, \mnras, 464, 3369

\bibitem[{{Zavala} {et~al.}(2023){Zavala}, {Buat}, {Casey}, {Finkelstein}, {Burgarella}, {Bagley}, {Ciesla}, {Daddi}, {Dickinson}, {Ferguson}, {Franco}, {Jim{\'e}nez-Andrade}, {Kartaltepe}, {Koekemoer}, {Le Bail}, {Murphy}, {Papovich}, {Tacchella}, {Wilkins}, {Aretxaga}, {Behroozi}, {Champagne}, {Fontana}, {Giavalisco}, {Grazian}, {Grogin}, {Kewley}, {Kocevski}, {Kirkpatrick}, {Lotz}, {Pentericci}, {P{\'e}rez-Gonz{\'a}lez}, {Pirzkal}, {Ravindranath}, {Somerville}, {Trump}, {Yang}, {Yung}, {Almaini}, {Amor{\'\i}n}, {Annunziatella}, {Arrabal Haro}, {Backhaus}, {Barro}, {Bell}, {Bhatawdekar}, {Bisigello}, {Buitrago}, {Calabr{\`o}}, {Castellano}, {Ch{\'a}vez Ortiz}, {Chworowsky}, {Cleri}, {Cohen}, {Cole}, {Cooke}, {Cooper}, {Cooray}, {Costantin}, {Cox}, {Croton}, {Dav{\'e}}, {de La Vega}, {Dekel}, {Elbaz}, {Estrada-Carpenter}, {Fern{\'a}ndez}, {Finkelstein}, {Freundlich}, {Fujimoto}, {Garc{\'\i}a-Argum{\'a}nez}, {Gardner}, {Gawiser}, {G{\'o}mez-Guijarro}, {Guo}, {Hamilton}, {Hathi}, {Holwerda}, {Hirschmann},
  {Huertas-Company}, {Hutchison}, {Iyer}, {Jaskot}, {Jha}, {Jogee}, {Juneau}, {Jung}, {Kassin}, {Kurczynski}, {Larson}, {Leung}, {Long}, {Lucas}, {Magnelli}, {Mantha}, {Matharu}, {McGrath}, {McIntosh}, {Medrano}, {Merlin}, {Mobasher}, {Morales}, {Newman}, {Nicholls}, {Pandya}, {Rafelski}, {Ronayne}, {Rose}, {Ryan}, {Santini}, {Seill{\'e}}, {Shah}, {Shen}, {Simons}, {Snyder}, {Stanway}, {Straughn}, {Teplitz}, {Vanderhoof}, {Vega-Ferrero}, {Wang}, {Weiner}, {Willmer}, {Wuyts}, \& {Ceers Team}}]{Zavala2023}
{Zavala}, J.~A., {Buat}, V., {Casey}, C.~M., {et~al.} 2023, \apjl, 943, L9

\bibitem[{{Zavala} {et~al.}(2021){Zavala}, {Casey}, {Manning}, {Aravena}, {Bethermin}, {Caputi}, {Clements}, {Cunha}, {Drew}, {Finkelstein}, {Fujimoto}, {Hayward}, {Hodge}, {Kartaltepe}, {Knudsen}, {Koekemoer}, {Long}, {Magdis}, {Man}, {Popping}, {Sanders}, {Scoville}, {Sheth}, {Staguhn}, {Toft}, {Treister}, {Vieira}, \& {Yun}}]{Zavala2021}
{Zavala}, J.~A., {Casey}, C.~M., {Manning}, S.~M., {et~al.} 2021, \apj, 909, 165

\bibitem[{{Zhang} \& {Hao}(2018)}]{Zhang2018}
{Zhang}, K. \& {Hao}, L. 2018, \apj, 856, 171

\end{thebibliography}
%

\begin{appendix}
\section{Images and photometric fluxes}\label{sec:images}
We report in Fig. \ref{fig:cutouts} the CEERS-14821 cutouts in the available HST and JWST filters, all converted to a pixel scale of 30 mas/pixel. The source is completely undetected at $\lambda\leq1.6\,\mu m$, barely detected in the F200W $(S/N<2$) and robustly detected ($S/N>5$) in all filters at $\lambda\geq2.7\,\mu m$. This would classify this source as a F200W-dropout. CEERS-14821 is also very compact, as it is barely resolved in the F277W and F356W filters and unresolved in the F410M and F444W ones. This is visible in Fig. \ref{fig:profiles}, where we report the light profiles and their respective models.\\

In Table \ref{tab:fluxes} we report the total fluxes measured for CEERS-14821 in the available HST and JWST filters. We highlight that the HST/F140W is particularly affected by large noise, as is also visible from the cutouts.

\begin{figure}
    \centering
    \includegraphics[width=\linewidth]{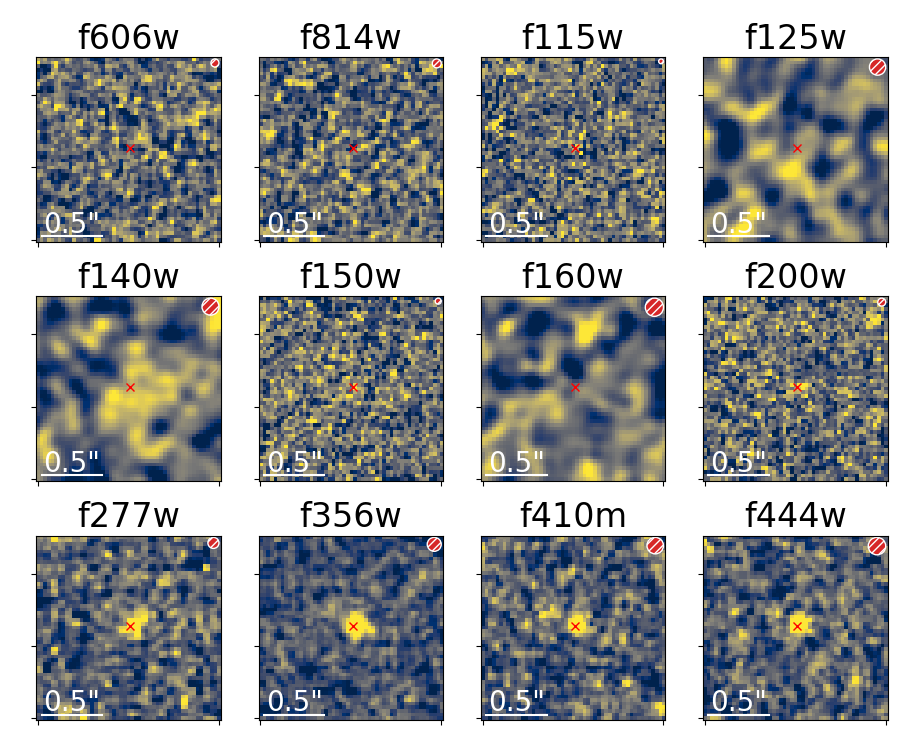}
    \caption{CEERS-14821 HST and JWST cutouts converted to a pixel scale of 30 mas/pixel. The position of the source is marked with a red cross. We report the scale in arcseconds in the bottom left and the filter PSF (diameter equal to the full-width-half-maximum) in the upper right.}
    \label{fig:cutouts}
\end{figure}

 \begin{figure}
     \centering
     \includegraphics[width=0.45\linewidth]{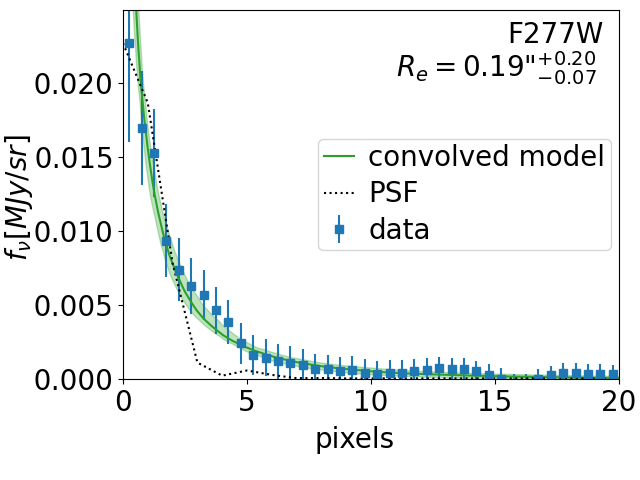}
     \includegraphics[width=0.45\linewidth]{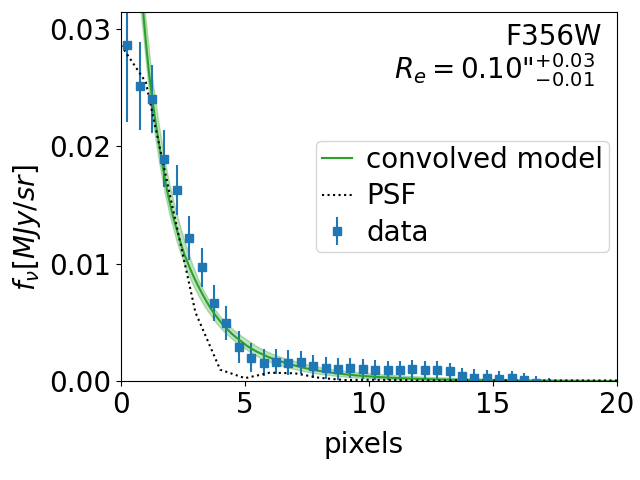}
     \includegraphics[width=0.45\linewidth]{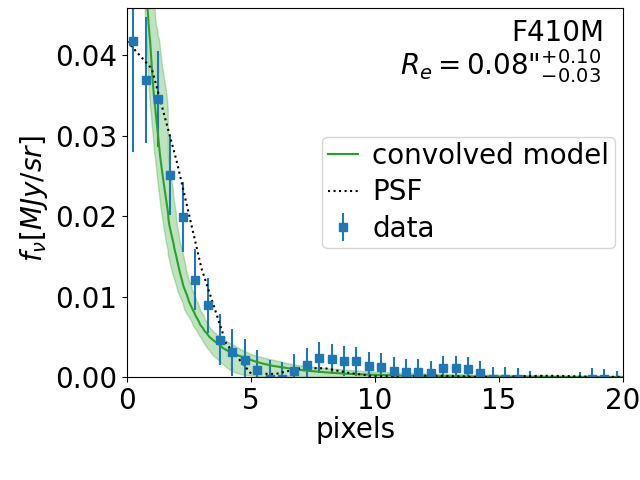}
     \includegraphics[width=0.45\linewidth]{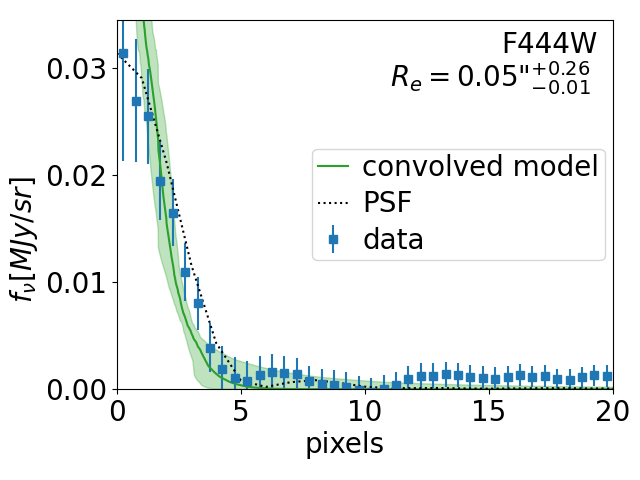}
     \caption{Light profile of CEERS-14821 in the four filters where it is detected. We show the measure profile (blue squares), the PSF normalised to the central pixel (dotted black line) and the convolved model (solid green line).}
     \label{fig:profiles}
 \end{figure}

\begin{table}[]
     \caption{Measured photometry of CEERS-14821.}
     \centering
     \begin{tabular}{cc|cc}
       filter   &  $f_{nJy}$ [nJy] &  filter   &  $f_{nJy}$ [nJy]\\
\hline
HST/F606W &$ -0.45 \pm 4.34 $ & HST/F160W & $ 7.05 \pm 6.49 $ \\
HST/F814W & $ 6.41 \pm 5.78 $ & JWST/F200W & $ 4.77 \pm 2.87 $ \\
JWST/F115W & $ -0.34 \pm 3.37 $ & JWST/F277W & $ 16.59 \pm 2.92 $ \\
HST/F125W & $ 1.83 \pm 7.61 $ & JWST/F356W & $ 31.64 \pm 2.79 $ \\
HST/F140W$^{a}$ & $ 26.03 \pm 13.07 $ & JWST/F410M & $ 36.05 \pm 5.76 $ \\
JWST/F150W & $ 6.17 \pm 3.95 $ & JWST/F444W & $ 27.62 \pm 3.96 $ \\

     \end{tabular}
     \label{tab:fluxes}
     \tablefoot{$^{a}$ As visible from the cutouts shown in Fig. \ref{fig:cutouts}, F140W is affected by particularly high noise.}
 \end{table}
\section{Line fit}\label{sec:lfmore}

\begin{figure*}
    \centering
    \includegraphics[width=\linewidth]{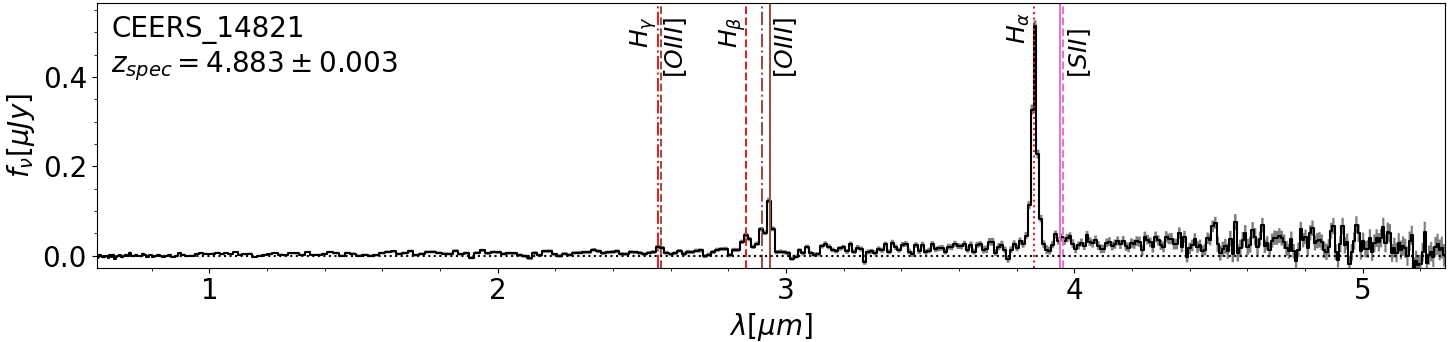}
    \caption{Observed NIRSpec/PRISM spectrum of CEERS-14821. The vertical lines highlight the visible nebular emission lines.}
    \label{fig:spectrum}
\end{figure*}

In this appendix we give more information on the continuum subtraction and the line fit applied to derive the line fluxes. The observed spectrum is shown in Fig. \ref{fig:spectrum}.

First, we applied a multiplicative factor that takes into account wavelength-dependent slit loss. This is derived by fitting a second-order polynomial function to the ratio of the spectrum, convolved with the broadband filters, and the photometric fluxes. 

Then, we removed the stellar continuum. Given its low S/N, we decided to consider a simple approach and fit it with a quadratic function, after masking the wavelengths affected by strong nebular emission lines. This simple approach could result in an underestimation of the $H_{\beta}$ and $H_{\gamma}$ line fluxes, given that we are not considering possible stellar absorption features. Based on the evolutionary stellar synthesis models by \citet{GonzalezDelgado2005}, the $H_{\beta}$ absorption feature can have an equivalent width up to 10\AA, but given that it strongly depends on stellar age and CEERS-14821 is going through a burst of star formation (see Sect. \ref{sec:discussion}), the contribution is probably smaller. Even in the worst situation, this would underestimate the $H_{\beta}$ flux by less than $10\%$ and overestimate the $A_{V}$ by less than 0.4. 

We described each nebular emission line using a Gaussian function. Line fitting was performed in the rest frame, using the redshift derived from the [$\ion{\rm O}{III}$]\,5007\AA line, that is, $z_{\mathrm{spec}}=4.883\pm0.003$, which is consistent with the redshift reported by \citet{deGraaff2024}. Given the low spectral resolution of the PRISM, we left the line centre free to vary within one resolution element (i.e. $10\AA$ for $H_{\beta}$ and $8\AA$ for $H_{\alpha}$). The fit was performed simultaneously for [$\ion{\rm O}{III}$]\,5007\AA, [$\ion{\rm O}{III}$]\,4959\AA, and $H_{\beta}$, as well as for $H_{\alpha}$, [$\ion{\rm N}{II}$]\,6548\AA, and [$\ion{\rm N}{II}$]\,6584\AA, due to line blending. In the multicomponent fit, we assumed that the width of the lines is the same and the ratios [$\ion{\rm O}{III}$]\,4959\AA/[$\ion{\rm O}{III}$]\,5007\AA and [$\ion{\rm N}{II}$]\,6584\AA/[$\ion{\rm N}{II}$]\,6548\AA are equal to the theoretical value of one-third. We repeat the fit 1000 times, randomising the flux considering a normal distribution with standard deviation equal to the flux uncertainties. 
All lines are unresolved or barely resolved. We tried a multicomponent fit also for the $[\ion{S}{II}]6717,6731\AA$ doublet, but the S/N is too low, therefore we obtain only an upper limit for the combined line fluxes. The $H_{\gamma}$ line could also be contaminated by the $[\ion{O}{III}]4363\AA$ line, but given the overall low S/N we do not try to deblend them.

In Fig. \ref{fig:linefit} we report the multi-component fit of the \ha/, \niia/, and \niib/ as well as the fit of the \hb/, \oiiia/, and \oiiib/. As visible from the residuals, the fits are consistent with the data within the flux uncertainties. However, while the necessity for three components in the second case is obvious, the line blending is more severe in the $H_{\alpha}+[\ion{\rm N}{II}]$ complex. Using the Kolmogorov-Smirnov test, the difference among the residuals considering a single Gaussian component and three components is not significant. The difference in the reduced $\chi^{2}$ is $\Delta \chi^{2}=0.3$, with the multi-component fit showing the lowest $\chi^{2}$. Looking instead at the Bayesian information criterion (BIC), the multicomponent fit is preferred to the single-component fit, since for the single Gaussian fit we obtain $\rm BIC=71.0$, while for the multicomponent fit we derived a lower value of $\rm BIC=63.4$. Overall, the improvement in the fit when considering the \niia/ and \niib/ lines is present, but its statistical importance is limited. 

In Fig. \ref{fig:hanii_deg} we show the fluxes of \ha/ and \niib/ derived by repeating the multicomponent fit 1000 times and randomly evaluating the observed fluxes within the uncertainties each time. As expected, an anticorrelation is present, where the \niib / flux increases with decreasing \ha / flux. We took this degeneracy into account when deriving the gas-phase metallicity using these nebular emission lines, but this result needs to be taken with caution. If we instead consider a single-component fit, given the reduced improvement when including the \niia/ and \niib/ lines, we obtain a total flux of $f_{H_{\alpha}}=(116.93\pm2.50)\times10^{-19}\,erg/s/cm^{2}$. This would increase even more the dust-extinction derived from the Balmer decrement, up to $A_V=3.1\pm0.4$ with \citetalias{Calzetti2000} and $A_V=4.8_{-0.7}^{+0.6}$ with \citetalias{CF00}. However, we would also discard the estimate of the gas-phase metallicity based on the \niib/ line. Overall, spectroscopic data at higher spectral resolution are necessary to properly estimate dust extinction and gas-phase metallicity, but CEERS-14821 still has a large dust extinction, even if we consider a single-Gaussian component to describe the $H_{\alpha}+[\ion{N}{II}]$ complex.

In Table \ref{tab:lines} we report the measured fluxes and upper limits of the nebular emission lines.
\begin{figure}
    \centering
    \includegraphics[width=\linewidth]{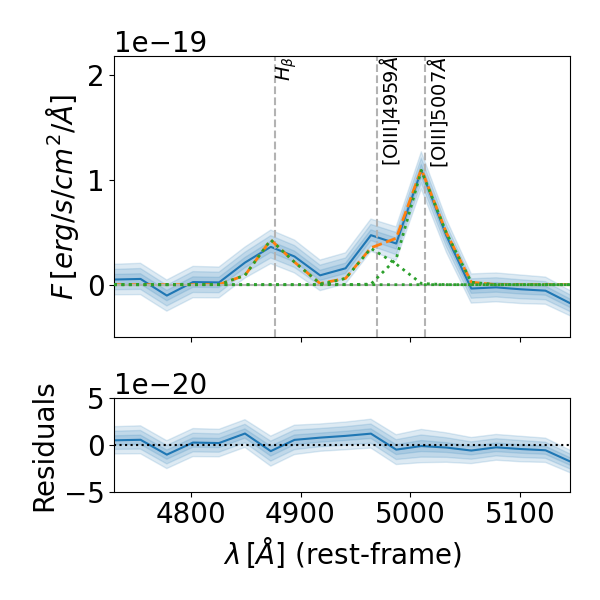}
    \includegraphics[width=\linewidth]{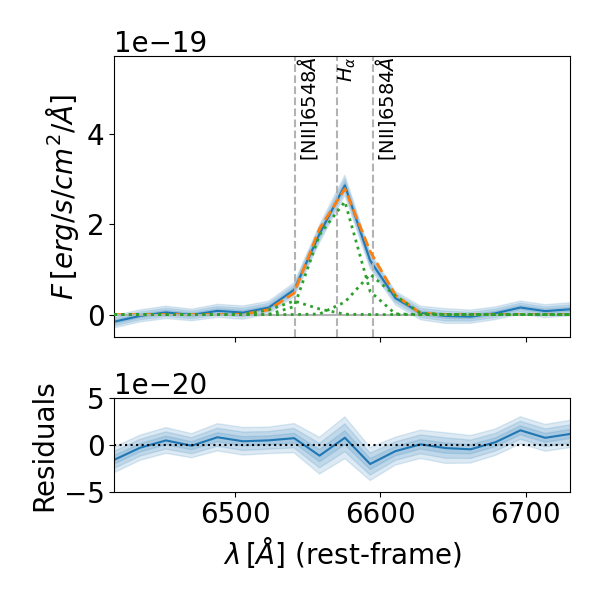}
    \caption{Multi-component fit of the $H_{\beta}$, \oiiib/ and \oiiia/ lines (top) and the $H_{\alpha}$, \niia/ and \niib/ lines (bottom). The solid~blue lines show the observed continuum-subtracted spectrum, with the blue shaded areas indicating the $1\sigma$, $2\sigma$, and $3\sigma$ uncertainties. Dotted green lines show the different components of the fit, while the orange dashed lines show the total model. Grey vertical lines show the expected center of the nebular emission lines. The bottom panels show the residuals of the fit. The spectral resolution is around $R=100$ near \hb/ and $R=200$ near \ha/.}
    \label{fig:linefit}
\end{figure}

\begin{figure}
    \centering
    \includegraphics[width=\linewidth]{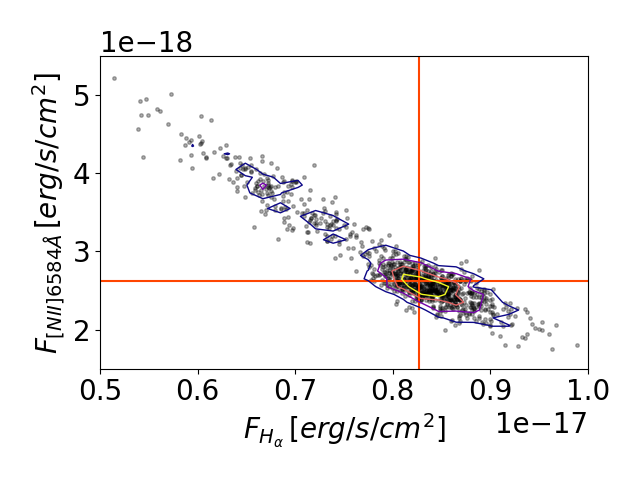}
    \caption{\ha/ and \niib/ flux estimates, derived by applying a multi-component fit to 1000 different randomisations of the observed fluxes. Solid red lines indicate the median values reported in Table \ref{tab:lines}, while the coloured contours show the 20\%, 40\%, 60\%, and 80\% of the distribution.}
    \label{fig:hanii_deg}
\end{figure}

\begin{table}[]
    \caption{Measured line fluxes.}
    \centering
    \begin{tabular}{c|c}

    line & $F\, [erg/s/cm^{2}]$ \\
    \hline
    $[\ion{O}{II}]3727\AA$ & $<8.92\times 10^{-19}\,^{a}$\\
    \hg/ + [\ion{O}{III}]4363\AA & $(7.91\pm2.57)\times 10^{-19}$ \\
    \heii/ & $<8.03\times 10^{-19}\,^{a}$\\
    \hb/ & $(15.32\pm1.85)\times 10^{-19}$ \\
    \oiiib/ & $(13.63\pm0.64)\times 10^{-19}$ \\
    \oiiia/ & $(40.89\pm1.94)\times 10^{-19}$ \\
    $[\ion{O}{I}]6300\AA$ & $<6.67 \times 10^{-19}\,^{a}$\\
    \ha/$^{b}$ & $(83.08\pm8.29)\times 10^{-19}$ \\
    \niia/$^{b}$ & $(8.77\pm2.03)\times 10^{-19}$ \\
    \niib/$^{b}$ & $(26.32\pm7.09)\times 10^{-19}$ \\
    $[\ion{S}{II}]6717,6731\AA$ & $<14.44\times 10^{-19}\,^{a}$ \\
    \end{tabular}
    \label{tab:lines}
    \tablefoot{$^{a}$ This is a $5\sigma$ upper limit, derived considering the local flux uncertainties and a full width at half maximum equal to two times the spectral resolution. $^{b}$ These lines are severely blended. See Appendix \ref{sec:lfmore} for an in-depth analysis. $^{b}$ this line estimation may be contaminated by the $[\ion{O}{III}]4363\AA$ line.}
\end{table}

\section{Impact of the reddening law slope}\label{sec:slope}

Previous studies have observed a flattening of the attenuation curve with increasing $A_{V}$ \citep[e.g.][]{Chevallard2013,Salmon2016,Lofaro2017,Boquien2022,Seille2022}, which would impact the physical properties derived. We therefore derived again the dust extinction based on the observed \ha//\hb/ line ratio, but assuming the dust extinction by \citepalias{CF00} with the slope free to vary. As visible in Fig. \ref{fig:Av_slopes}, even considering extremely steep dust-extinction curves (i.e. $\delta_{BC}=-2.$) we would measure $A_{V}>1$, which is still too large for a galaxy with the stellar mass of CEERS-14821. Based on the study by \citet{Boquien2022}, which however includes galaxies with stellar masses above $10^{9.25}\,M_{\odot}$, an overestimation in the dust-extinction slope would result in an overestimation of the stellar mass up to 0.3 dex, slightly increasing the tension with the literature stellar mass vs. dust extinction relation. In contrast, if we consider a flat dust extinction curve, the stellar mass could be underestimated by 0.3 dex \citep{Boquien2022}, but at the same time the dust extinction rapidly diverges to values above $A_{V}=4$. Therefore, even if we change the slope of the dust extinction, CEERS-14821 remains an outlier in the $M_{*}-A_{V}$ relation.

\begin{figure}
    \centering
    \includegraphics[width=\linewidth]{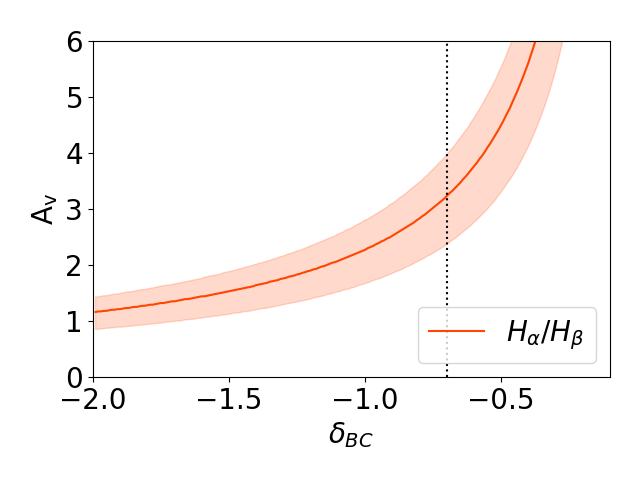}
    \caption{Dust extinction values derived for different slopes of the dust extinction law, considering the observed \ha//\hb/ line ratio. The vertical dotted line shows the value of $\delta_{BC}=-0.7$. }
    \label{fig:Av_slopes}
\end{figure}

\label{sec:SEDmore}
\section{Spectro-photometric SED fit}\label{sec:SEDmore}
In this Appendix, we give some additional information on the SED fit.

\subsection{BAGPIPES}
When fitting the spectro-photometric data with BAGPIPES \citep{Carnall2018}, we considered the 2016 version of the \citet{Bruzual2003} single stellar population models, leaving the metallicity free to vary from $5\%$ solar to 2.5 times solar. As mentioned in the main text, for dust extinction, we considered the reddening law by \citetalias{Calzetti2000} or \citetalias{CF00}, using a Gaussian prior centred on the value derived from the spectrum divided by a factor of 2.2 and 3.4, respectively, to convert from nebular to continuum dust extinction. Removing the prior on the dust extinction has no impact on the physical properties when assuming the \citetalias{CF00}  dust extinction law, while produce a mild decrease in $A_{V}$ and SFR when considering the \citetalias{Calzetti2000} dust extinction law.
We included nebular continuum and emission lines derived from a pre-computed model grid considering an ionisation parameter (i.e., the dimensionless ratio of densities of ionising photons to hydrogen) ranging from $\rm log_{10}(U)=-4$ to $-1$. We let the stellar masses be free to vary between $10^{6}$ and $10^{12.5}\,M_{\odot}$ considering a flat, uninformative prior. Finally, we considered a non-parametric star-formation history \citep{Leja2019}, considering three bins of 10, 20 and 70 Myr and then five other bins, equally distributed in logarithmic space, between 100 Myr and the age of the Universe (i.e., 1180 Myr). We assumed two different prior for the SFH, a continuity prior and bursty one \citep{Leja2019,Tacchella2022}. The complete list of parameters and priors is listed in Table \ref{tab:paramSED}.

\begin{table*}[]
    \caption{Parameters and priors of the BAGPIPES spectro-photometric fit.}
    \centering
    \begin{tabular}{cccc}
        Parameter & range & prior & hyper-parameters \\
        \hline
        stellar mass & ($10^{6},10^{12.5}$) $M_{\odot}$ & logarithmic & \\
        stellar metallicity & (0.05,2.5) $Z_{\odot}$ & uniform & \\
        ionisation parameter log(U) & ($10^{-4},10^{-1}$) & logarithmic & \\
        \multirow{2}{*}{log[d(SFR)]} & \multirow{2}{*}{(-10,10)} & Student’s t - continuity & $\mu=2,\,\rm \nu=0.3$\\
         &  & Student’s t - bursty & $\mu=2,\,\rm \nu=1.0$\\
        \hline
        \multicolumn{4}{c}{\citetalias{Calzetti2000} reddening law} \\
        $A_{V,cont}$  & (0,6) & Gaussian & $\mu=0.97,\,\sigma=0.26$ \\
        $A_{V,neb}/A_{V,cont}$ & 2.27 & fixed & \\
        \hline
        \multicolumn{4}{c}{\citetalias{CF00} reddening law} \\
        $A_{V,cont}$ & (0,6) & Gaussian & $\mu=1.1,\,\sigma=0.63$ \\
        $A_{V,neb}/A_{V,cont}$ & 3.0 & fixed & \\
        reddening law slope $n$ & -0.7 & fixed & \\
    \end{tabular}
    \label{tab:paramSED}
\end{table*}

In Figure \ref{fig:SEDfit} we show the comparison between the best SED models, considering both dust extinction laws and the available spectro-photometric data, while in Figure \ref{fig:sfh} we show the output SFH. In Figures \ref{fig:corner_c00} and \ref{fig:corner_cf00} we instead show the corner plots related to the two fits. These figures show that CEERS-14821 is going through a burst of star formation, as most of the star-formation happens in the last 10Myr. In particular, $SFR (<10Myr)=16_{-2}^{+2}\,M_{\odot}/yr$ and $SFR (<10Myr)=48_{-6}^{+7}\,M_{\odot}/yr$, depending on the assumed reddening laws, while in the other age bins $\leq0.1\,M_{\odot}/yr$. Overall, the SFR is lower than the values derived from the dust-corrected \ha/ line, but still consistent within 2$\sigma$.

\begin{figure}
    \centering
    \includegraphics[width=\linewidth]{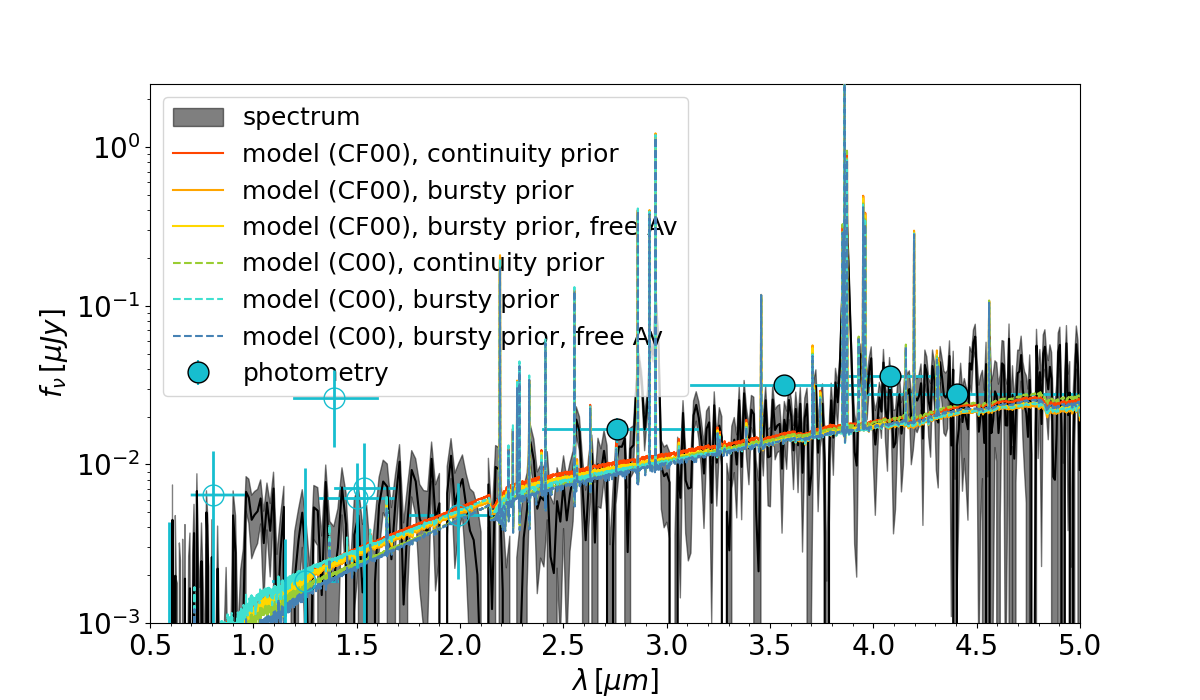}
    \caption{Spectro-photometric SED fit performed with BAGPIPES. We show the observed spectrum (black solid lines), the observed photometry (cyan circles) with empty symbols indicating filters with $S/N<3$, and the SED models obtained assuming the reddening law by \citetalias[][solid red, orange, and yellow lines]{CF00}, and \citetalias[][dashed green, cyan, and blue lines]{Calzetti2000}. In both cases we show the models derived assuming different priors on SFH and $A_{v}$.}
    \label{fig:SEDfit}
\end{figure}

\begin{figure}
    \centering
    \includegraphics[width=\linewidth]{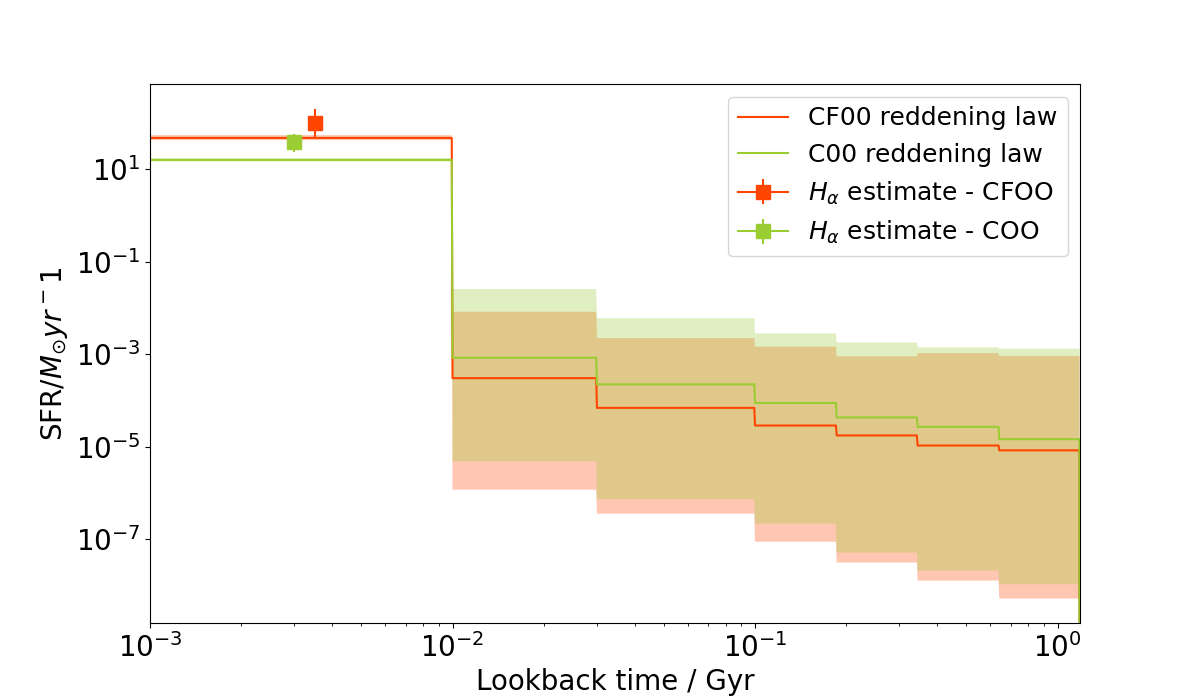}
    \caption{The SFH derived from the spectro-photometric SED fitting assuming the reddening law by \citetalias[][solid red, orange, and yellow lines]{CF00}, and \citetalias[][dashed green, cyan, and blue lines]{Calzetti2000}.}
    \label{fig:sfh}
\end{figure}

\begin{figure*}
    \centering
    \includegraphics[width=\linewidth]{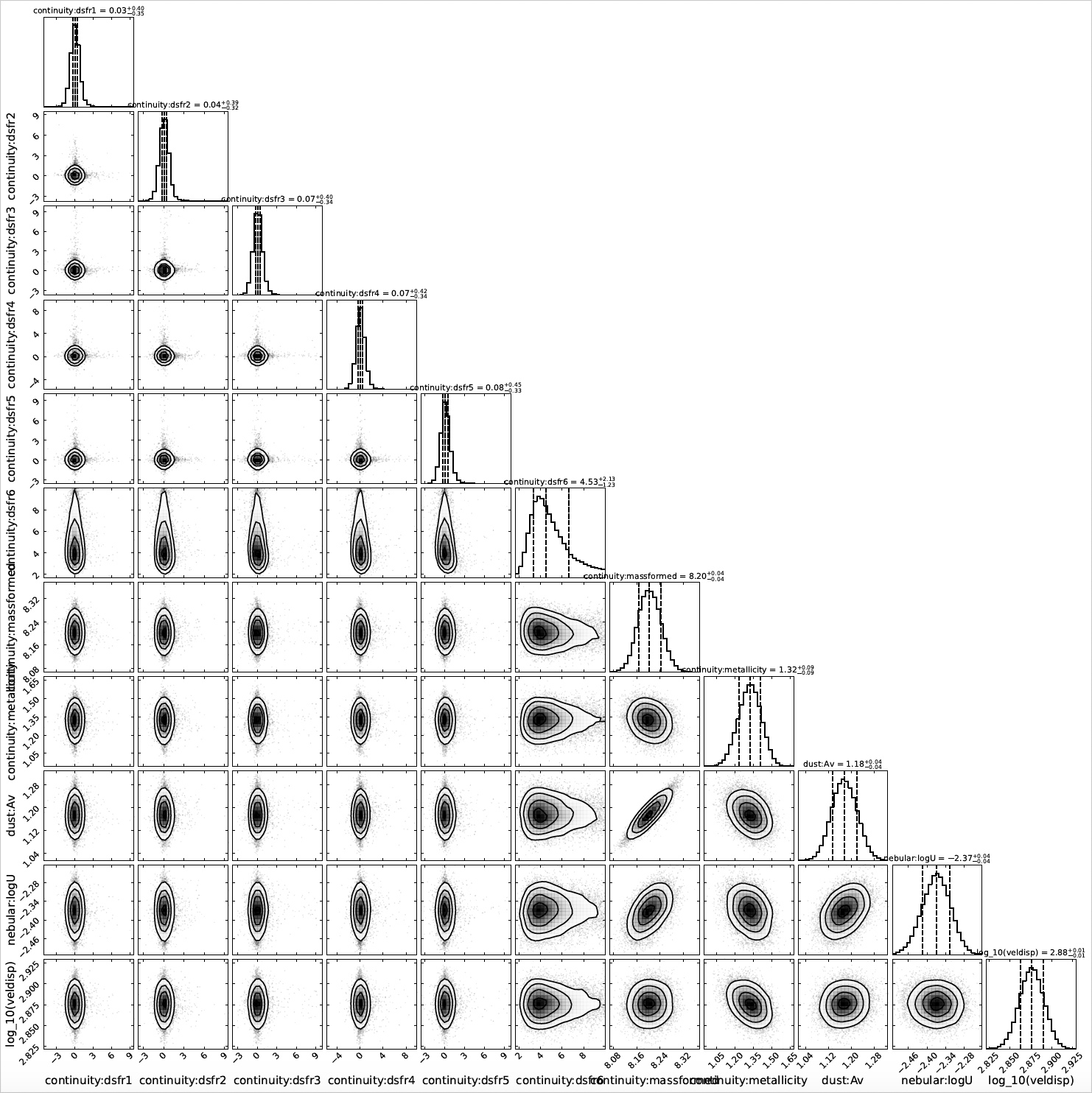}
    \caption{Corner plot of the fit considering the reddening law by \citetalias{Calzetti2000}, the continuity prior on the SFH and the $A_{v}$ prior. In this case, stellar mass does not include mass loss. $\rm dSFR_{i}$ indicates the SFR in the $i$ age bin with decreasing redshift.}
    \label{fig:corner_c00}
\end{figure*}

\begin{figure*}
    \centering
    \includegraphics[width=\linewidth]{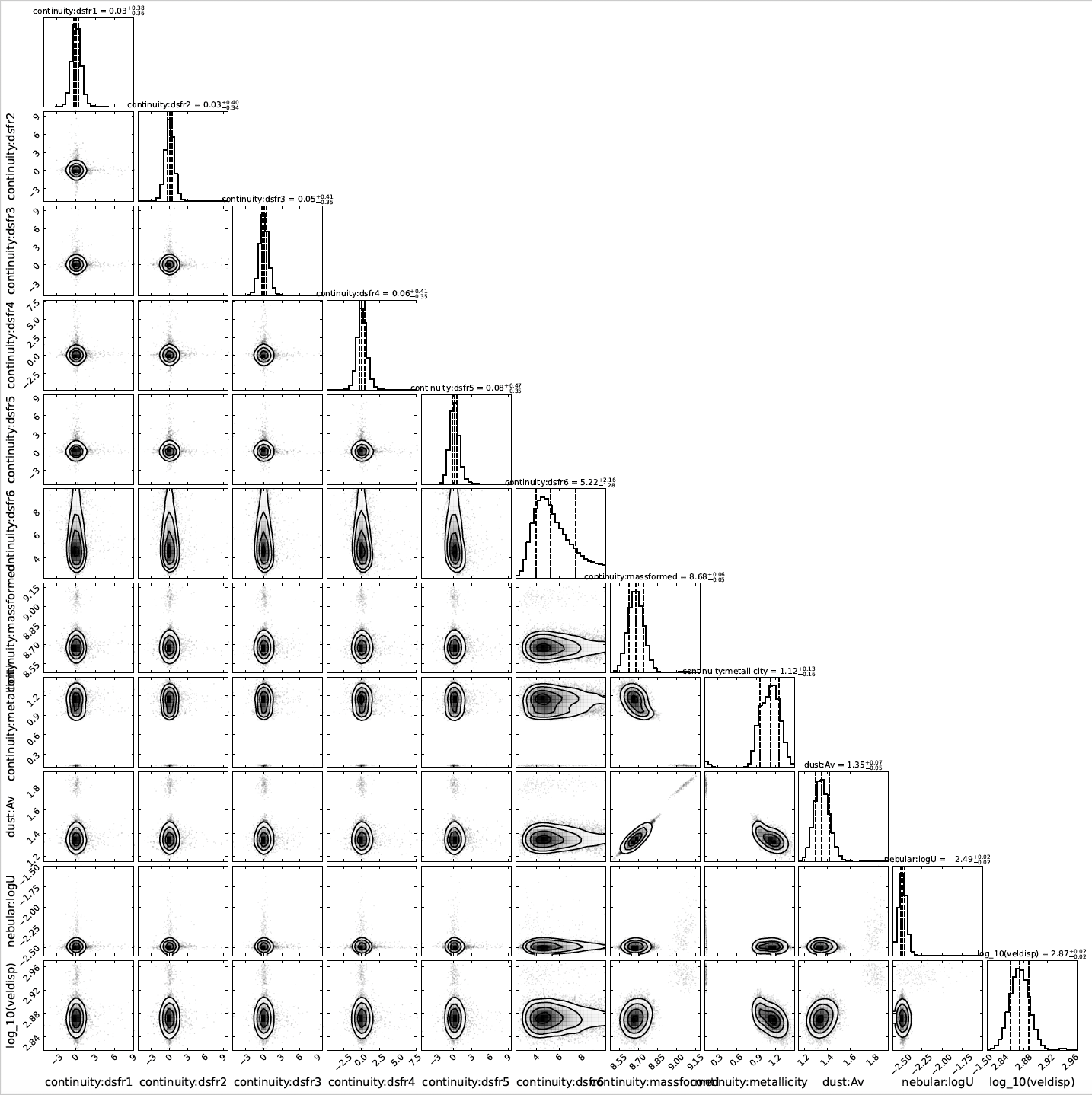}
    \caption{As Figure \ref{fig:corner_c00}, but considering the reddening law by \citetalias{CF00}. }
    \label{fig:corner_cf00}
\end{figure*}

\subsection{\texttt{CIGALE}}\label{sec:cigale}

We investigate the possible presence of an AGN using the photometric information, given that the diagnostics based on line ratios are not conclusive, using the code \texttt{CIGALE} \citep{Boquien2019}. The current version of BAGPIPES does not include an AGN component. In the fit, which is based only on photometric data because of code limitations, we set the redshift to the spectroscopic value. We considered a double exponential SFH with the addition of a possible extra burst 20Myr-old, which is motivated by the results based on BAGPIPES with a non-parametric SFH, and a \citetalias{CF00} dust extinction law with a slope of -0.7. The complete list of the free parameters considered is listed in Table \ref{tab:cigalegrid}, while the comparisons between the best SED model and the available photometric data are shown in Figure \ref{fig:SEDfit_cigale}.

We obtained an AGN fraction of $f_{AGN}=0.22\pm0.27$, derived with respect to the total dust luminosity, which does not exclude an AGN component, but is consistent with a negligible impact. Due to the large number of free parameters necessary to fit the AGN component, the stellar mass is less constrained than in the previous fit done with BAGPIPES, resulting in $\rm log_{10}(M/M_{\odot})=9.64\pm0.33$. The dust attenuation is large, $\rm A_V=2.8\pm1.3$, which is consistent with CEERS-14821 still being in the HELM selection (see Fig. \ref{fig:MAV}). In the text we consider as fiducial the BAGPIPES runs, given the limited AGN contribution and because they fit the spectrum together with the photometry and they are based on a non-parametric SFH.

\begin{table*}
    \centering
    \caption{Grid utilised for the free parameters of our \texttt{CIGALE} SED-fitting run.}
    \begin{tabular}{p{9.5cm} p{3.5cm} p{4cm}} 
        \texttt{CIGALE} fit parameters & Grid values & Description \\
        \cmidrule(lr){1-3}
        \multicolumn{3}{l}{Double exponential SFH [\texttt{sfh2exp} module]} \\
        $\tau_\text{main}$ & 2000, 6000, 10000 & e-folding time of the main stellar population model in Myr \\
        Age & 1000, 2000, 5000, 10000, 13000 & Age of the main stellar population in the galaxy in Myr \\
        $\tau_{burst}$ & 5., 10., 25., 50.0, 100. & e-folding time of the late starburst population model in Myr \\
        $f_{burst}$ & 0.1, 0.25, 0.5, 0.75 & Mass fraction of the late burst population \\
        \cmidrule(lr){1-3}
        \multicolumn{3}{l}{SSP component [\texttt{bc03} module]} \\
        Z & 0.0001, 0.004, 0.008, 0.02, 0.05 & Metallicity\\
        \cmidrule(lr){1-3}
        \multicolumn{3}{l}{Nebular component [\texttt{nebular} module]} \\
        $\log \text{U}$ & -4, -3, -2 & Logarithmic ionization parameter \\
        \cmidrule(lr){1-3}
        \multicolumn{3}{l}{Dust attenuation component [\texttt{dustatt\_modified\_CF00} module]} \\
        $A_\text{V, ISM}$ & 0.5, 1, 1.5, 2.5, 4, 5, 6 & V-band attenuation in the interstellar medium\\
        \cmidrule(lr){1-3}
        \multicolumn{3}{l}{AGN component [\texttt{fritz2006} module]} \\
        $\beta$ & -1, -0.5, 0 & Dust density distribution parameter \\
        $\gamma$ & 0, 2, 4 & Dust density distribution parameter \\
        $\Psi$ & 0.001, 50.1, 80.1 & Angle between the equatorial axis and line of sight \\
        $\text{f}_\text{AGN}$ & 0., 0.1, 0.25, 0.5, 0.75 & AGN fraction with respect to the total dust luminosity \\
        \cmidrule(lr){1-3}
        \cmidrule(lr){1-3}
        \label{tab:cigalegrid}
    \end{tabular}
    \tablefoot{In this table we show exclusively the parameters we allowed to vary in our analysis. All the remaining parameters are fixed and their value coincides with the default ones assigned by \texttt{CIGALE}.}
\end{table*}

\begin{figure}
    \centering
    \includegraphics[width=\linewidth]{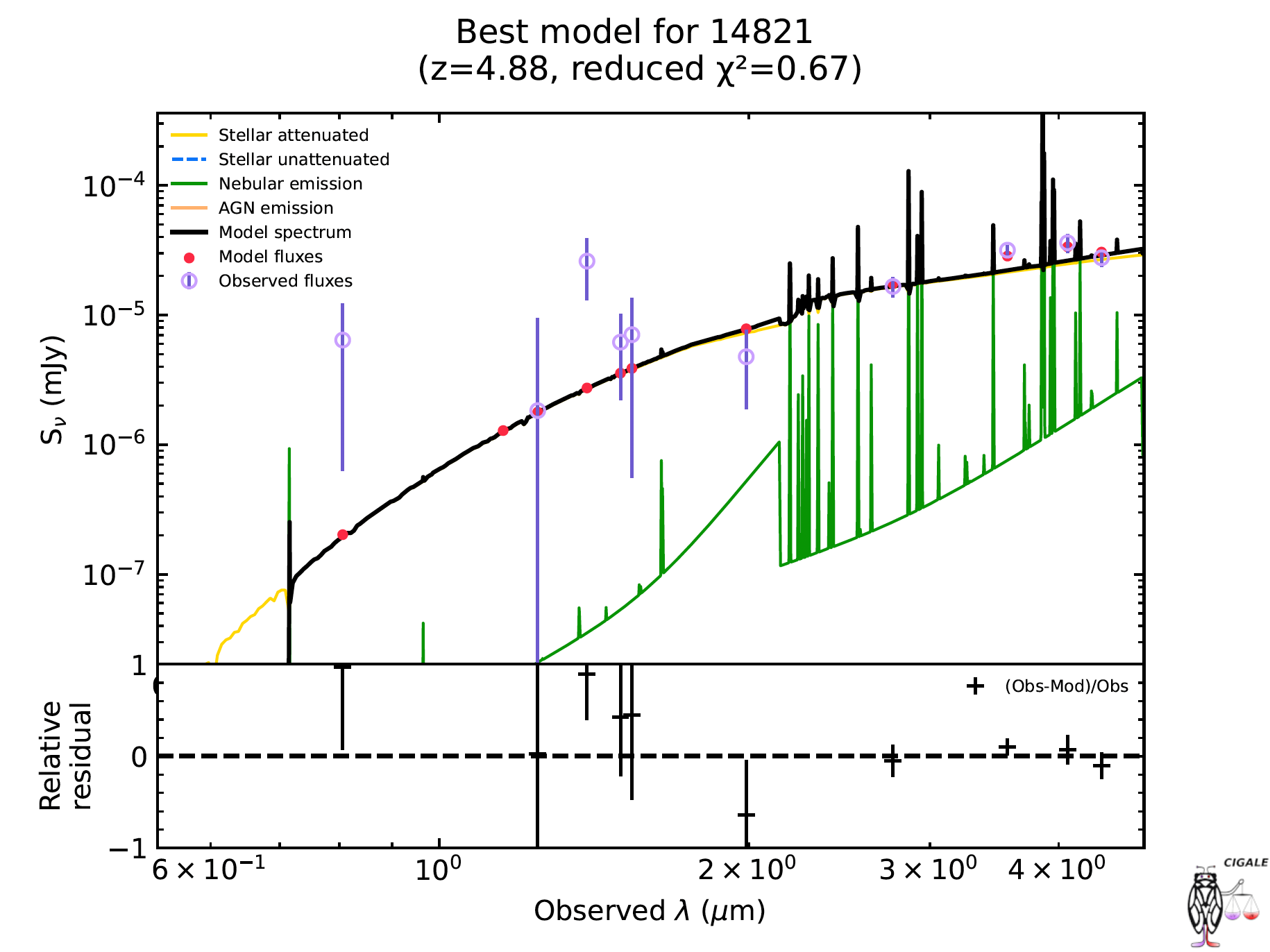}
    \caption{Photometric SED fit performed with \texttt{CIGALE}. We report the observed photometry (purple open circles), the model SED (solid black line) with the different components (coloured lines). The AGN emission (orange line) is so negligible to be too faint to be visible in the plot.}
    \label{fig:SEDfit_cigale}
\end{figure}

\end{appendix}
\end{nolinenumbers}
\end{document}